\documentclass[journal, onecolumn, 12pt]{IEEEtran}
\renewcommand{\baselinestretch}{2.0}
\usepackage{cite}
\usepackage{ifpdf}
\usepackage{array}
\ifpdf
\usepackage{graphicx}
\usepackage{subfigure}
\usepackage[svgnames]{xcolor}
\else
\usepackage[dvipdfmx]{graphicx}
\usepackage[dvipdfmx,svgnames]{xcolor}
\fi
\usepackage{amsmath,amssymb,amscd}
\usepackage{verbatim}
\usepackage{enumerate}

\newtheorem{rem}{Remark}
\newtheorem{prop}{Proposition}
\usepackage{algorithm}
\usepackage{algpseudocode}
\usepackage{tikz}
\usepackage{epstopdf}
\begin{document}
\renewcommand{\baselinestretch}{1.55}
\title{Control-Data Separation with Decentralized Edge \\ Control in Fog-Assisted Uplink Communications}

\author{\large Jinkyu Kang, Osvaldo Simeone, Joonhyuk Kang and Shlomo Shamai (Shitz)
\thanks{This work was presented in part at IEEE wireless communications and networking conference (WCNC), San Francisco, CA, USA, Mar. 2017 \cite{Kang17WCNC}. The work of O. Simeone was partially supported by the U.S. NSF through grant CCF-1525629. O. Simeone has also received funding from the European Research Council (ERC) under the European Union’s Horizon 2020 research and innovation programme (grant agreement No 725731). The work of S. Shamai has been supported by the European Union's Horizon 2020 Research And Innovation Programme, grant agreement no. 694630.

Jinkyu Kang is with the School of Engineering and Applied Sciences (SEAS), Harvard University, Cambridge, MA 02138, USA (Email: jkkang@g.harvard.edu).

Osvaldo Simeone is with the Department of Informatics, King's College London, London, UK (Email: osvaldo.simeone@kcl.ac.uk). 

Joonhyuk Kang is with the Department of Electrical Engineering, Korea Advanced Institute of Science and Technology (KAIST) Daejeon, South Korea (Email: jhkang@ee.kaist.ac.kr).

Shlomo Shamai (Shitz) is with the Department of Electrical Engineering, Technion, Haifa, 32000, Israel (Email: sshlomo@ee.technion.ac.il).
}
}
\maketitle
\renewcommand{\baselinestretch}{2.0}
\begin{abstract}
Fog-aided network architectures for 5G systems encompass wireless edge nodes, referred to as remote radio systems (RRSs), as well as remote cloud center (RCC) processors, which are connected to the RRSs via a fronthaul access network. RRSs and RCC are operated via Network Functions Virtualization (NFV), enabling a flexible split of network functionalities that adapts to network parameters such as fronthaul latency and capacity. This work focuses on uplink communications and investigates the cloud-edge allocation of two important network functions, namely the control functionality of rate selection and the data-plane function of decoding. Three functional splits are considered: {\it{(i)}} Distributed Radio Access Network (D-RAN), in which both functions are implemented in a decentralized way at the RRSs; {\it{(ii)}} Cloud RAN (C-RAN), in which instead both functions are carried out centrally at the RCC; and {\it{(iii)}} a new functional split, referred to as Fog RAN (F-RAN), with separate decentralized edge control and centralized cloud data processing. The model under study consists of a time-varying uplink channel in which the RCC has global but delayed channel state information (CSI) due to fronthaul latency, while the RRSs have local but more timely CSI. Using the adaptive sum-rate as the performance criterion, it is concluded that the F-RAN architecture can provide significant gains in the presence of user mobility. 
\end{abstract}

\section{Introduction} \label{Sec:Intro}
The evolution of the wireless network architecture traces a line from the decentralized implementation of control and data functionalities in conventional Distributed Radio Access Network (D-RAN) through the centralization of the protocol stack in Cloud-RAN (C-RAN) \cite{ChinaMobile13, ChinaMobileNFGI} to the more recent fog-aided proposals with flexible functional splits between cloud and edge nodes \cite{5GNORMA}. An important motivation for the latest shift to fog-aided solutions is the realization that a fully centralized C-RAN system entails significant, and possibly prohibitive, requirements on the fronthaul connections between edge nodes and cloud, see, e.g., \cite{Fettweis14SPMAG, Simeone16JCN} and references therein. Furthermore, the development of the Network Functions Virtualization (NFV) technology makes adaptive cloud-edge functional splits realizable via software \cite{Rost17arXiv}.

The demarcation line between the functionalities to be implemented at the cloud and at the edge is typically drawn to include a given number of physical-layer functions at the edge nodes, such as synchronization, FFT/IFFT and resource demapping \cite{Fettweis14SPMAG, Chang16ICC}. The body of work concerned with edge-cloud functional splits generally aims at assessing the trade-off between performance and fronthaul capacity overhead of different demarcation lines. 

In light of these developments, references \cite{Dotsch13Bell, Rost2014WCL, Khalili16TETT} explore the application of the \emph{data-control separation architecture} \cite{Tafazolli15CST} as the guiding principle underlying the separation of functionalities between edge and cloud with the aim of addressing fronthaul latency limitations. Specifically, \cite{Dotsch13Bell} puts forth the idea of performing the control decisions of the uplink hybrid automatic repeat request (HARQ) protocol at an edge node, while keeping the computationally expensive operation of data decoding at the cloud processor. As shown in \cite{Rost2014WCL, Khalili16TETT, Gulati16VTC}, this approach can yield significant reductions in transmission latency thanks to the capability of the edge nodes to provide quick feedback to the mobile users with limited fronthaul overhead. 

An important lesson learned from \cite{Dotsch13Bell, Rost2014WCL, Khalili16TETT, Gulati16VTC} is that the implementation of some control functionalities at the edge can be an enabler for the reduction of transmission latency even in the presence of significant delays on the fronthaul links. A work that provides related insights in the different set-up of a multi-hop network with orthogonal links is \cite{Johnston15ISIT}. Reference \cite{Johnston15ISIT} shows that centralized scheduling based on delayed channel state information (CSI) can be outperformed by local scheduling decisions, as long as each network node has more current CSI of its incoming and outgoing links with respect to the centralized scheduler.

\begin{figure}[t]
\centering
\subfigure{\label{Fig:D-RAN}\includegraphics[height=4cm]{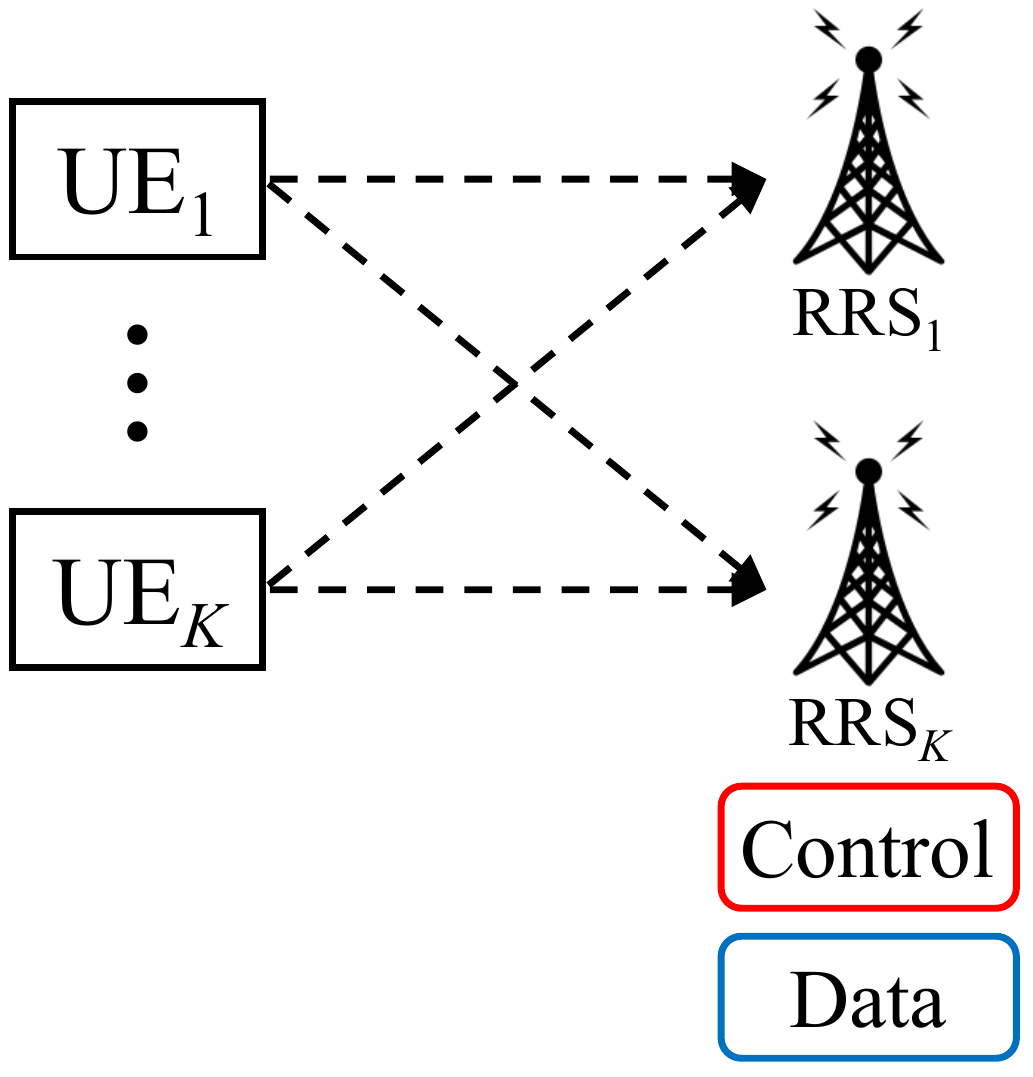}} \hspace{0.05cm} \rule[0ex]{1pt}{4cm} \hspace{0.05cm}
\subfigure{\label{Fig:C-RAN}\includegraphics[height=4cm]{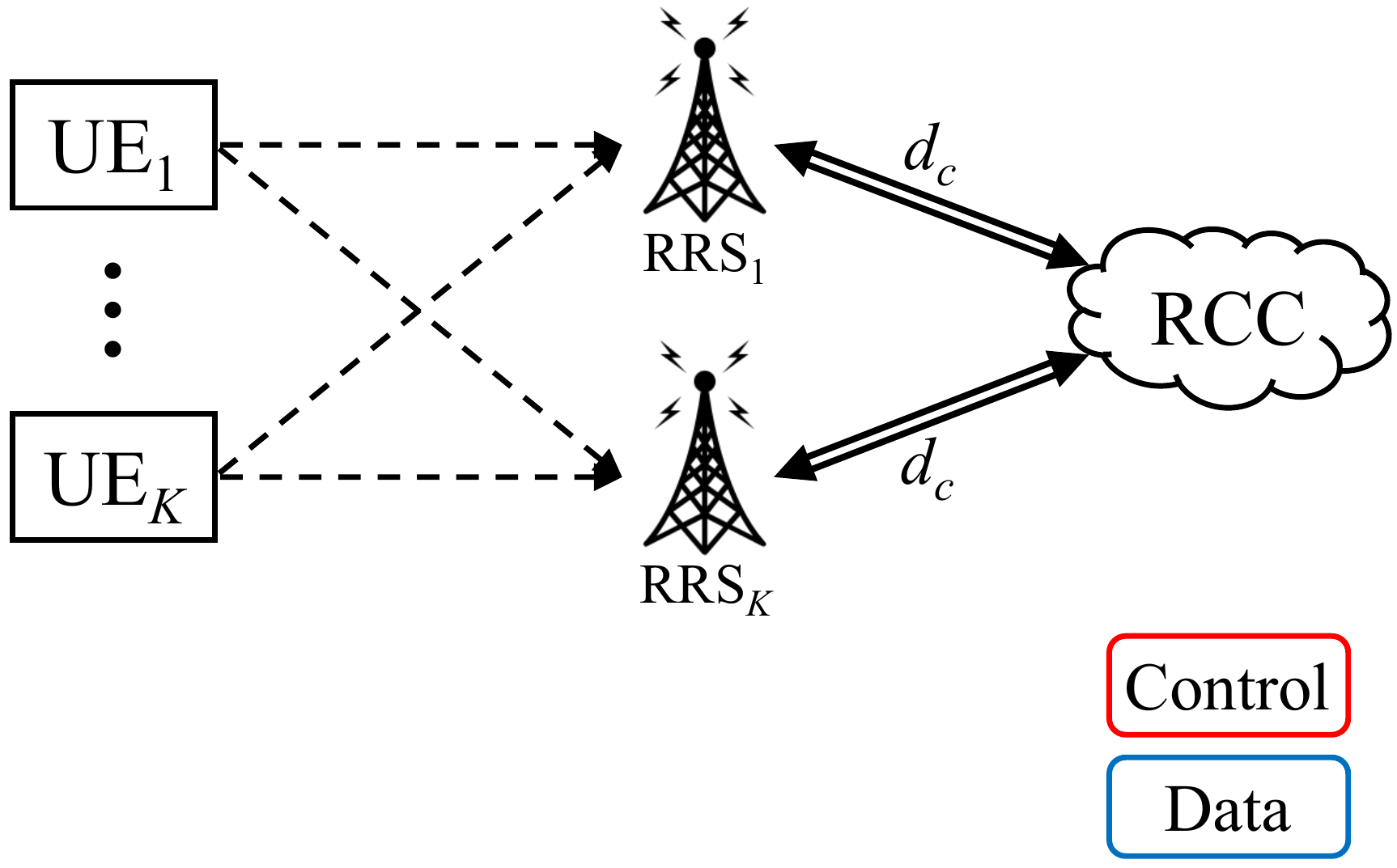}} \hspace{0.05cm} \rule[0ex]{1pt}{4cm} \hspace{0.05cm}
\subfigure{\label{Fig:F-RAN}\includegraphics[height=4cm]{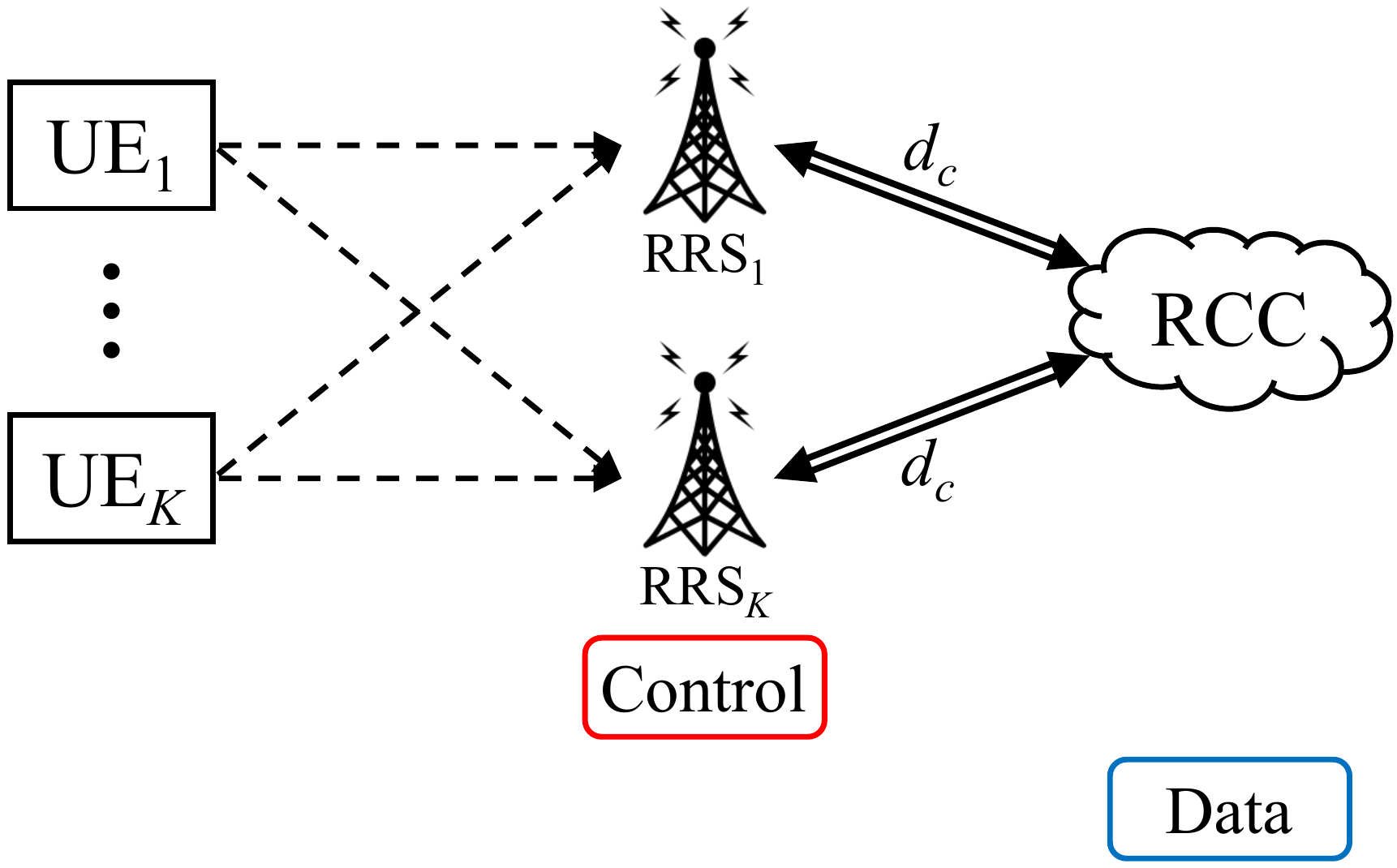}} \\
\vspace{-0.5cm}
{\footnotesize \hspace{0.6cm} {(a) D-RAN \hspace{4.15cm} (b) C-RAN \hspace{5.5cm} (c) F-RAN} \hspace{1.8cm}}
\vspace{-0.6cm}
\caption{Three radio access network architectures for different control-data functional splits between cloud and edge: (a) D-RAN, (b) C-RAN, and (c) F-RAN.}
\label{Fig:Three-RAN}
\end{figure}

In this work, as illustrated in Fig. \ref{Fig:Three-RAN}, we study the optimal functional split of control and data plane functionalities at the edge nodes, referred to as remote radio systems (RRSs) \cite{ChinaMobileNFGI}, and at the cloud, referred to as remote cloud center (RCC) \cite{ChinaMobileNFGI}, for uplink communication. We specifically focus on the following functionalities: {\it{(i)}} the {\it{control plane}} functionality of the data rate selection, and {\it{(ii)}} the {\it{data plane}} functionality of data decoding. We aim at assessing the impact of fronthaul latency on the relative performance of different splits, whereby rate selection and data decoding may be performed separately at either cloud or edge. 

As summarized in Fig. \ref{Fig:Three-RAN}, we specifically consider three functional splits: {\it{(i)}} D-RAN, in which both rate selection and data decoding are implemented at each edge; {\it{(ii)}} C-RAN, whereby both rate selection and data decoding are instantiated at cloud; and {\it{(iii)}} Fog-RAN (F-RAN), whereby the control function of rate selection is performed at the edge, while data decoding is implemented at cloud. The latter functional split is studied here for the first time. The approach is motivated by the idea discussed above of leveraging decentralized control to counteract fronthaul delays. We remark that the label ``F-RAN'' has been used in works such as \cite{Sengupta16arXiv} to indicate systems with decentralized caching at the RRSs and centralized processing at the RCC. Here we suggest to use the term more generally to describe fog-based solutions involving both cloud and edge operations.

As seen in Fig. \ref{Fig:Fog-assistedsystem}, the model under study consists of a time-varying uplink channel in which the RCC processor has global but delayed CSI due to fronthaul latency, while the RRSs have local CSI with a lower delay. Using the adaptive sum-rate as the performance criterion (see, e.g., \cite{SreekumarTIT15}), the mentioned functional splits based on the control-data separation architecture are compared through analysis and numerical results. 

The rest of the paper is organized as follows. We describe the system model and performance metric in Sec. \ref{Sec:SM}. We analyze the three radio access network architectures in Fig. \ref{Fig:Three-RAN} for different control-data functional splits between RCC and RRSs: D-RAN in Sec. \ref{Sec:D-RAN}, C-RAN in Sec. \ref{Sec:C-RAN}, and F-RAN in Sec. \ref{Sec:F-RAN}. In Sec. \ref{Sec:Numerical}, numerical results are presented. Concluding remarks are summarized in Sec. \ref{Sec:Conclusion}.

\section{System Model and Performance Metric} \label{Sec:SM}
\begin{figure}[t]
\centering
\vspace{-0.5cm}
\includegraphics[width=12cm]{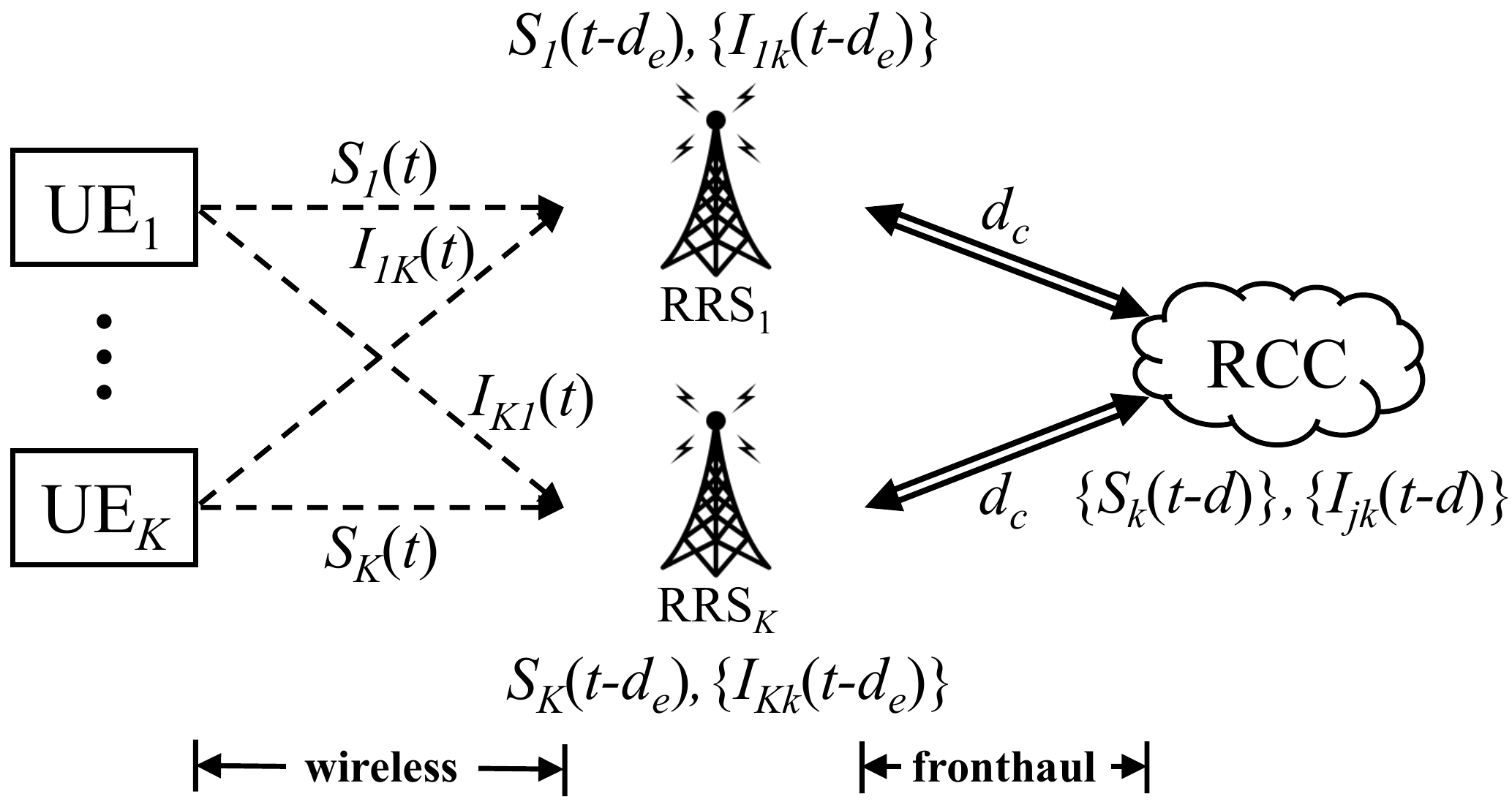}
\vspace{-0.5cm}
\caption{Uplink of the considered fog-assisted system. Communication on the fronthaul links entails a two-way latency of $d_c$ time slots (of the wireless interface), while the time elapsed between CSI measurement at the edge and uplink scheduling is $d_e$ time slots. The figure indicates the CSI available when selecting the rates for transmitting in slot $t$ at the RRSs for D-RAN and F-RAN and at the RCC for C-RAN. Note that the CSI available for decoding can be assumed to be timely since it can be estimated from pilots included in the data frame, as seen in Fig. \ref{Fig:Protocol timeline}.}
\label{Fig:Fog-assistedsystem}
\end{figure}

We consider the uplink of a fog-assisted system illustrated in Fig. \ref{Fig:Fog-assistedsystem}, which consists of $K$ remote radio systems (RRSs), a remote cloud center (RCC), and $K$ active user equipments (UEs). We assume that user-cell association has been carried out, so that each UE $i$ is associated to a given RRS $i$, and we have the same number of active UEs and RRSs. We denote the set of all UEs and RRSs as $\mathcal{K} = \{1, \dots, K\}$. As further detailed below and illustrated in Fig. \ref{Fig:Three-RAN}, we consider three different cloud-edge splits, namely: {\it{(i)}} D-RAN: The RCC is not present and both rate selection and data decoding for UE $k$ are carried out at RRS $k$; {\it{(ii)}} C-RAN: The RCC implements both rate selection and data decoding for all UEs; {\it{(iii)}} F-RAN: In this novel functional split, the RRS $k$ performs rate selection for UE $k$ while data decoding for all UEs is performed at the RCC.

\begin{figure}[t]
\centering
\includegraphics[width=12cm]{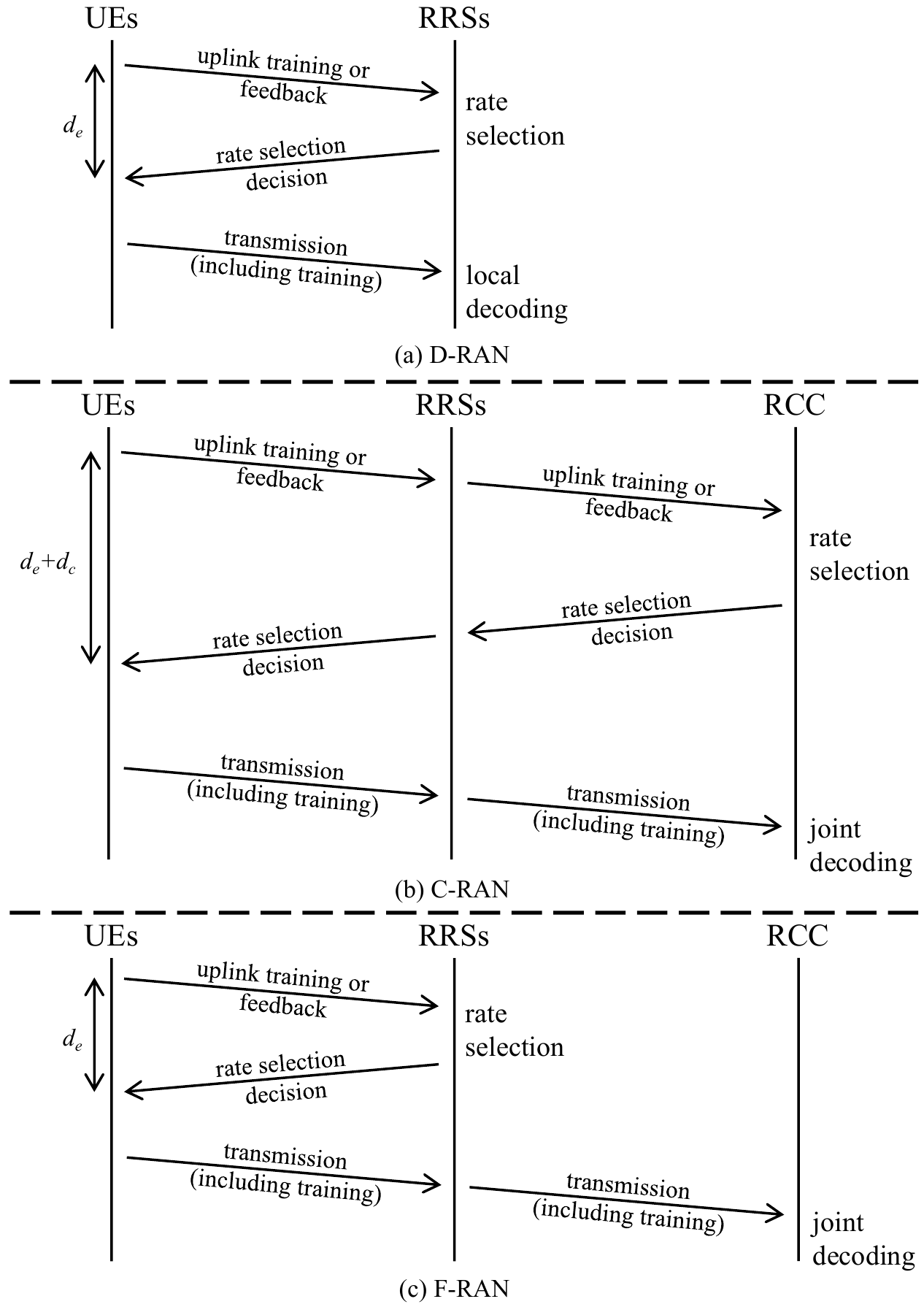}
\vspace{-0.5cm}
\caption{Protocol timeline (time increasing from top to bottom) for: (a) D-RAN; (b) C-RAN; and (c) F-RAN. For D-RAN (a) and F-RAN (c), the latency between uplink training transmission or CSI feedback and the time slot allocated for uplink transmission equals $d_e$ time slots of the wireless channel. For C-RAN (b), in addition to the scheduling delay $d_e$, one needs to consider the latency associated with two-way fronthaul communication between RRSs and RCC, which equals $d_c$ time slots.}
\label{Fig:Protocol timeline}
\end{figure}

An important role in the analysis is played by the timeliness of the CSI available at the RRS and RCC at the time of rate selection. In particular, as illustrated in Fig. \ref{Fig:Protocol timeline}, for D-RAN and F-RAN, we assume that the latency between uplink training transmission or CSI feedback and the time slot allocated for uplink transmission equals $d_e$ time slots of the wireless channel. As an example, the latency contributions for uplink transmission in LTE Release 14 \cite{LTERel14Latency} are Scheduling Request (SR) periodicity, uplink scheduling delay and uplink grant transmission. For a transmission time interval (TTI) of, say, $0.5$$-$$1$ ms, the latency $d_e$ can be large as $1$$-$$2$ slots \cite{LTERel14Latency}. For C-RAN, in addition to the delay $d_e$, one needs to consider the two-way communication between RRSs and RCC on the fronthaul. This entails a latency equal to $d_c$ time slots. The fronthaul transport latency is reported to be around $0.25$ ms in \cite{NGMNOnline} for single-hop fronthaul links and can amount to multiple milliseconds in the presence of multihop fronthauling, while fronthaul-related processing at the RCC can take fractions to a few milliseconds \cite{Nikaein15CONF}. As a result, for TTI of $0.5$$-$$1$ ms, the RCC CSI latency $d_c$ can be as large as $3$$-$$5$ slots. 

\subsection{Channel Model}
At each time slot $t$, the instantaneous power received at RRS $i$ from UE $i$ is denoted as $S_i(t)$, while the received power for the cross-channel between an UE $i$ and the RRS $j \neq i$ is denoted as $I_{ji}(t)$. These are assumed to vary across the time index $t=1,2, \dots, T$, which runs over the transmission intervals, according to a Markov model. This model can be obtained, for instance, by approximating the standard Clarke's model via quantization, see, e.g., \cite{Wang95TVT}. The channel matrix between the UEs and the RRSs at any channel use $k$ of the transmission interval $t=1,2, \dots, T$ can be written as 
\begin{equation} \label{CH_Model}
{\bf{H}} (t,k) = [{\bf{h}}_1^T (t,k), \dots, {\bf{h}}_{K}^T (t,k)]^T, 
\end{equation}
with ${\bf{h}}_j (t,k) = [\sqrt{I_{j1}(t)} e^{j \theta_{j1} (t,k)}, \dots, \sqrt{S_j(t)} e^{j \theta_{jj} (t,k)},  \dots, \sqrt{I_{jK}(t)} e^{j \theta_{jK} (t,k)} ]$, where the phases $\theta_{ji} (t,k)$ are uniformly distributed in the interval $[0, 2 \pi)$, mutually independent as per the standard Rayleigh fading model, and vary in an ergodic manner over the channel use index $k$ within each transmission interval $t$. 

\begin{figure}[t!]
\centering
\includegraphics[width=12cm]{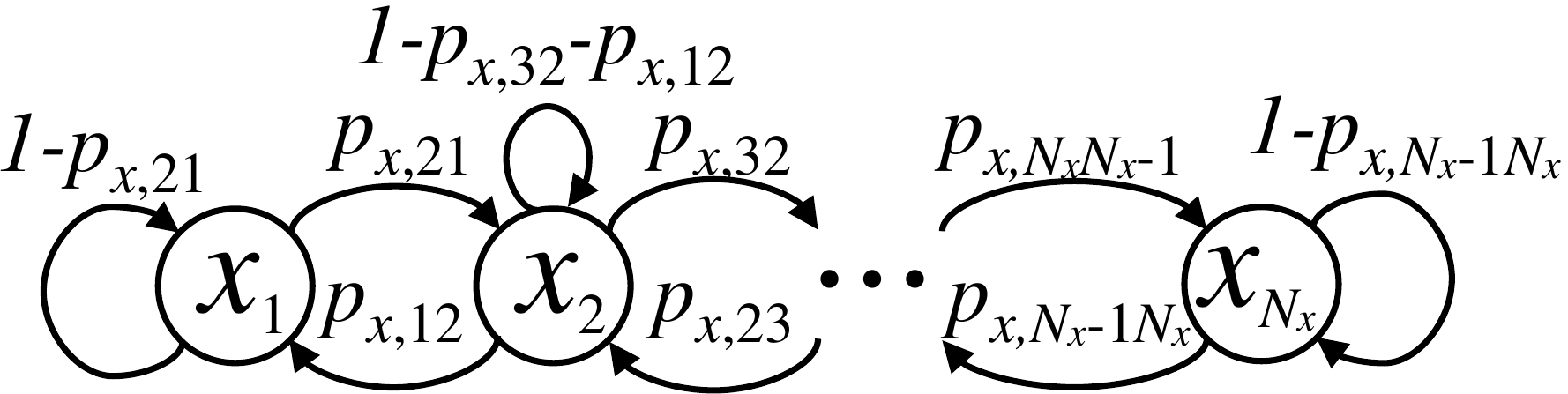}
\caption{Markov model for the direct channel and cross-channel instantaneous power processes with $N_x$ states for $x \in \{S, I\}$ \cite{Wang95TVT}.}
\label{Fig:MarkovModel for general case}
\end{figure}

As illustrated in Fig. \ref{Fig:MarkovModel for general case}, the direct instantaneous fading power $S_i(t)$ can take $N_S$ values, indexed in ascending order as $\{S_1, \dots, S_{N_S}\}$, and is governed by a Markov chain with transition probabilities $p_{S, mn} = \textrm{Pr}[ S_i (t+1) = S_{m} | S_i(t) = S_{n} ]$. In a similar manner, the cross-channel power $I_{ji} (t)$ can take $N_I$ values, indexed in ascending order as $\{ I_{1}, \dots, I_{N_I}\}$, and varies according to a Markov chain with transition probabilities $p_{I, mn} = \textrm{Pr}[ I_{ji} (t+1) = I_{m} | I_{ji} (t) = I_{n} ]$. We denote the set of all states for the direct channel as $\mathcal{N}_S = \{S_1, \dots, S_{N_S}\}$ and for the cross-channel as $\mathcal{N}_I = \{I_1, \dots, I_{N_I}\}$. We recall that Markovian models are widely adopted for the evaluation of the performance of wireless systems (see, e.g., \cite{Wang95TVT, Wei10TVT, Zheng13TWC}). Note that, as in \cite{Wang95TVT}, channel variations can only occur between adjacent states, i.e., $p_{x,mn} = 0$ if $|m-n|>1$ for $x \in \{S, I\}$. Details on the quantization process from Clarke's model, which is assumed for the numerical results presented in Sec. \ref{Sec:Numerical}, can be found in Appendix \ref{Appendix: Finite state Markov channel}.

We conclude this subsection by introducing some useful notation. According to the adopted Markov model, the probability that the direct channel state changes from state $S_{n}$ to the state $S_{m}$, for $S_m, S_n \in \mathcal{N}_S$, after $d$ transmission intervals can be written as 
\begin{equation} \label{Prob_conditional}
{\textrm{Pr}} \left [ S_i (t) = S_{m} | S_i (t-d) = S_{n} \right ] = \beta_S^{m|n} (d),
\end{equation}
where the probability $\beta_S^{m|n} (d)$ is obtained as the $(m, n)$ entry of the matrix ${\bf{T}}_S^d$, with the transition matrix ${\bf{T}}_S$ having $p_{S, mn}$ as the $(m,n)$ entry, i.e., $[{\bf{T}}_S]_{m,n} = p_{S, mn}$. Moreover, the stationary probability $\pi_{S,m}$ for the state $S_{m}$ is obtained by solving the linear system as (see, e.g., \cite{MarkovChainBook})
\begin{equation} \label{Prob_SteadyState}
\pi_{S,m} = \sum_{S_n \in \mathcal{N}_S} \pi_{S,n} p_{S, mn},
\end{equation}
for $S_m \in \mathcal{N}_S$. Analogously, we define $\beta_I^{m|n} (d)$ as the $d$-step transition probability for the interference process, i.e., ${\textrm{Pr}} \left [ I_{ji} (t) = I_{m} | I_{ji} (t-d) = I_{n} \right ] = \beta_I^{m|n} (d)$ with $m, I_n \in \mathcal{N}_I$, and $\pi_{I,m}$ as the steady-state probability of the interference process, i.e., $\pi_{I,m} = \sum_{I_n \in \mathcal{N}_I} \pi_{I,n} p_{I, mn}$ for $I_m \in \mathcal{N}_I$. We also use the notation $\pi_{\bf{S}} = \prod_{m \in \mathcal{K}} \pi_{S,m}$ for the joint stationary probability of any given direct channel vector ${\bf{S}} \in \mathcal{N}_S^{K}$ and we also define $\pi_{\bf{I}} = \prod_{i \in \mathcal{K}} \prod_{j \in \mathcal{K}, j \neq i} \pi_{I, I_{ji}}$ for the joint stationary probability of any given cross channel vector ${\bf{I}} \in \mathcal{N}_S^{K-1 \times K}$. 
\subsection{Cloud-Edge Functional Splits} \label{Sec:SM;FuncSplit}
As discussed, we focus on the \textit{control} functionality of rate adaptation, or adaptive modulation and coding, that is, the selection of the transmission rates $R_j$ (bit/s/Hz) for any UE $j \in \mathcal{K}$, and on the \textit{data} plane functionality of data decoding. The three control-data functional splits under study (see Fig. \ref{Fig:Three-RAN}) are formalized below. \\
$\bullet$ {\textit{Distributed Radio Access Network}} (D-RAN): D-RAN amounts to the most conventional cellular implementation in which control and data plane functionalities are carried out at the RRSs, that is, at the edge. Accordingly, for each time slot $t$, each RRS $j$ selects rate $R_j$ for UE $j$ on the basis of {\it{local}} delayed CSI about the direct channel $S_j(t- d_e)$ and about the cross channel ${\bf{I}}_{j}(t- d_e)$ from all other UEs to the RRS $j$. This information can be obtained, e.g., by means of uplink training in a Time Division Duplex system or via feedback with Frequency Division Duplex (FDD). Moreover, each RRS $j$ individually performs decentralized local data decoding of the signal transmitted by UE $j$ by treating interference as noise. Since data packets are assumed to include training signals, we assume that channel decoding at each RRS can leverage current CSI about the data packet. \\
$\bullet$ {\textit{Cloud Radio Access Network}} (C-RAN): In the C-RAN architecture, the RCC carries out both control and data processing. Specifically, the RCC selects jointly all rates $\{R_j\}_{j \in \mathcal{K}}$ on the basis of {\it{global}} delayed CSI $\{ S_j(t- d) \}$ and $\{ {\bf{I}}_{j}(t- d) \}$ for $j \in \mathcal{K}$ about the channels from all UEs to the all RRSs. Note that the delay $d$ includes the additional fronthaul delay $d_c$ between RRSs and RCC as well as the scheduling delay $d_e$, i.e., $d = d_c + d_e$. Moreover, upon reception of the signals received by the RRSs on the fronthaul links, the RCC performs centralized joint data decoding. Again, CSI for date decoding can be estimated from the training sequences in the packet and hence timely CSI can be assumed for decoding. \\
$\bullet$ {\textit{Fog Radio Access Network}} (F-RAN): The novel F-RAN solution is a hybrid implementation with control processing at the edge and data processing at the cloud. In the proposed solution, each RRS $j$ selects the rate $R_j$ based on local delayed CSI as for D-RAN, while the RCC performs centralized joint data decoding on behalf of the RRSs as in C-RAN.
\subsection{Performance Metric}
To compare the different functional splits in Fig. \ref{Fig:Three-RAN}, we will use the performance metric of the {\textit{adaptive sum-rate}} (with no power control) used in \cite{SreekumarTIT15} and references therein. This is defined as the average sum-rate that can be achieved while guaranteeing no outage in each transmission slot. Note that an outage event corresponds to the case that the signal of {\it{at least}} one user is not decoded correctly. The average is taken here with respect to the steady-state distribution (\ref{Prob_SteadyState}) of the random channel gains $\{S_i (t)\}$ and $\{I_{ji}(t)\}$ for $i, j \in \mathcal{K}$ and $j \neq i$. To ensure that no outage occurs, in each transmission interval, transmission rates $\{R_j\}_{j \in \mathcal{K}}$ for all users are chosen by the RCC or by the RRSs, depending on the functional splits, so that successful decoding can be guaranteed. The adaptive sum-rate is the corresponding achievable average of the sum rates $\sum_{j=1}^{K} R_j$. 

More generally, we will consider the $\epsilon$-{\textit{outage adaptive sum-rate}}, which is defined as the maximum adaptive sum-rate under the constraints a (small) outage probability $\epsilon$ is allowed in each slot. We emphasize that an outage event is caused by the imperfect knowledge of the CSI at the time of rate selection. 

In the following sections, we analyze the performance in terms of the $\epsilon$-outage adaptive sum-rates of the mentioned control-data functional splits between RCC and RRSs in the presence of the scheduling delay $d_e$ and the fronthaul transmission delay $d_c$. 
\section{Distributed radio access network (D-RAN)} \label{Sec:D-RAN}
In this section, we study the conventional cellular implementation based on D-RAN. Accordingly, for each slot $t$, each RRS $j$ selects the transmission rate $R_j (t)$ for the user $j$ in its cell based on the available delayed direct channel $S_j (t-d_e)$ and cross-channels $I_{ji} (t-d_e)$ for $i \in \mathcal{K} \setminus \{j\}$. Furthermore, it performs local data decoding by treating interference from the out-of-cell user as noise. As a result, in a D-RAN, an outage event for the $j$-th RRS/UE pair occurs at time $t$ if the selected rate $R_j (t)$ is larger than the current available rate $C_j (S_j (t), \{ I_{ji} (t) \}) \triangleq \log_2 (1 + S_j (t)/(1+ \sum_{i=1, i \neq j}^{K} I_{ji} (t))$.

The adaptive outage sum-rate can then be expressed as a function of a conditional CDF of the achievable rates $C_j(S_j (t), \{I_{ji} (t)\})$ for each UE $j \in \mathcal{K}$, where the conditioning is over the delayed CSI $S_{j} (t-d_e)$ and $I_{ji}(t-d_e)$. This CDF is defined as
\begin{equation} \label{D-RAN:CDF}
F_{d_e} (R_j| S_j, {\bf{I}}_{j}) \triangleq \text{Pr} \left [ \left. C_j(S_j (t), \{I_{ji} (t)\}) < R_j \,\, \right | \,\, S_{j} (t-d_e) = S_j, I_{ji}(t-d_e) = I_{ji} \right ],
\end{equation}
where ${\bf{I}}_j = \{I_{ji}\}_{i \in \mathcal{K}}$ is the collection of the states for the cross-channels from all UEs to RRS. The conditional CDF (\ref{D-RAN:CDF}) can be computed in terms of the conditional probabilities $\beta_S^{m| S_j} (d_e)$ and $\beta_I^{m| I_{ji}} (d_e)$ as 
\begin{equation} \label{D-RAN:CDF2}
F_{d_e} (R_j| S_j, {\bf{I}}_{j}) = \sum_{ {\bf{I}}_{j}(t), S_j(t): \, C_j(t) < R_j} \beta_S^{S_j(t) | S_j} (d_e) \prod_{i \in \mathcal{K} \setminus \{j\}} \beta^{ I_{ji}(t) | I_{ji}}_I (d_e),
\end{equation}
where we have used the short-hand notation $C_j(t) = C_j(S_j (t), \{I_{ji} (t)\})$.

\prop \label{Prop:D-RAN}With D-RAN, an achievable $\epsilon$-outage adaptive sum-rate is given by 
\begin{equation} \label{AOSR:D-RAN}
R^{\text{D-RAN}} (d_e, \epsilon) =  \sum_{j \in \mathcal{K}} \mathbb{E}_{{\bf{S}}, {\bf{I}}} \left [ F^{-1}_{d_e} (\bar \epsilon | S_j, {\bf{I}}_j) \right],
\end{equation} 
where $F^{-1}_{d_e} (\bar \epsilon| S_j, {\bf{I}}_j)$ is the inverse of the conditional CDF (\ref{D-RAN:CDF2}), the average is taken with respect to the product distribution $\pi_{\bf{S}} \pi_{\bf{I}}$ and $\bar \epsilon = 1 - (1-\epsilon)^{1/K}$.
\begin{IEEEproof}
If each RRS $j$ chooses rate $R_j = F^{-1}_{d_e} (\bar \epsilon| S_j, {\bf{I}}_j)$, it is by construction guaranteed that, when $S_{j} (t-d_e) = S_j$ and $I_{ji}(t-d_e) = I_{ji}$, the individual probability of outage is no larger than $\bar \epsilon$. Since outage events of different users are independent, overall outage probability is no larger than $\epsilon = 1 - ( 1 - \bar \epsilon)^K$.
\end{IEEEproof}
\section{Cloud radio access network (C-RAN)} \label{Sec:C-RAN}
In a C-RAN, at any transmission interval $t$, the RCC performs rate adaptation in a centralized manner based on the available global and delayed CSI, namely $\{S_j (t-d)\}$ and $\{I_{ji}(t-d)\}$ for all $i, j \in \mathcal{K}$ with $i \neq j$, where the delay $d = d_e + d_c$ includes the edge and fronthaul delays. Furthermore, the RCC performs centralized joint data decoding on behalf of the connected RRSs. Given the complexity of the problem of analyzing the $\epsilon$-outage adaptive sum-rate for C-RAN, we first consider a simplified scenario with two RRS-UE pairs, in which the direct links have fixed fading power and the cross-channel have two states, i.e., $K=2$, $N_S=1$, and $N_I=2$. We then tackle the general case with multiple RRS-UE pairs and multiple channel states. 
\subsection{Analysis with two RRSs and UEs} \label{Sec:C-RAN;Special}
Here, we focus on a simplified scenario with two RRSs and UEs, namely $K=2$; fixed direct channels $S_i(t) = S$ for $i \in \mathcal{K}$, which may be realized in practice via power control; and cross-channels $I_{12}(t) \triangleq I_1 (t)$ and $I_{21}(t) \triangleq I_2 (t)$ taking values in a binary set $\mathcal{N}_I = \{I_L, I_H\}$ with $I_H \ge I_L$. Note that the latter assumption implies that the cross-channels can take either a ``low'' value $I_L$ or a ``high'' value $I_H$. To simplify the notation, we set the transition probabilities for the Markov chain describing the variation of the cross-channels as $p_{I, HL} \triangleq p$ and $p_{I, LH} \triangleq q$. Accordingly, the stationary probabilities for the ``low'' and ``high'' states of the cross-channels are obtained as
\begin{equation} \label{Prob_SteadyState}
\pi_L = \frac{q}{p+q} \,\,\,\,\, \text{and} \,\,\,\,\, \pi_H = \frac{p}{p+q},
\end{equation}
respectively.

To proceed, we define $C (I_1,I_2)$ as 
\begin{equation} \label{C-RAN:Capacity;two}
C (I_1,I_2) \triangleq \mathbb{E} [\log_2 \det ( {\bf{I}} + {\bf{H}} (I_1,I_2) {\bf{H}}^\dagger(I_1,I_2))],
\end{equation} 
where ${\bf{H}} (I_1,I_2) = [\sqrt{S} e^{j \theta_{11}} \,\, \sqrt{I_{1}} e^{j \theta_{12}}; \sqrt{I_{2}}e^{j \theta_{21}} \,\, \sqrt{S} e^{j \theta_{22}}]$. The expectation in (\ref{C-RAN:Capacity;two}) is taken over the random phases ${\pmb{\theta}} = [\theta_{11}, \theta_{12}, \theta_{21}, \theta_{22}]$, which are mutually independent and uniformly distributed in the interval $[0, 2 \pi]$. The quantity in (\ref{C-RAN:Capacity;two}) is the maximum achievable sum-rate in a time-slot with $I_1 (t) = I_1$ and $I_2 (t) = I_2$ if joint data decoding is performed at the RCC (see, e.g., \cite[Ch. 4]{GamalBook}). We will also use the notation $C_{xy}$ for $C (I_x, I_y)$ when $I_1 = I_x$ and $I_2 = I_y$ for $x, y \in \{L, H\}$. We finally observe that $C_{LH} = C_{HL}$.

\begin{figure}[t!] 
\centering
\subfigure[$C_{LH} \le C_{LL}/2 + \log_2(1+S+I_L)$]{\label{Fig:Capacity_region_aa}\includegraphics[height=5.5cm]{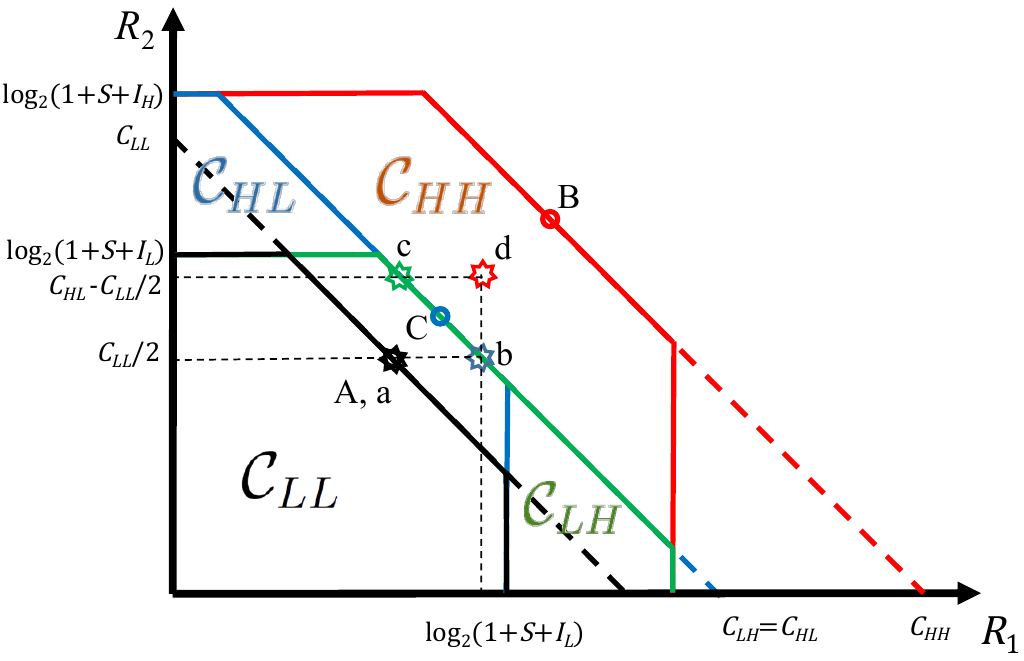}}
\subfigure[$C_{LL}/2 + \log_2(1+S+I_L) < C_{LH} \le 2 \log_2(1+S+I_L)$]{\label{Fig:Capacity_region_bb}\includegraphics[height=5.5cm]{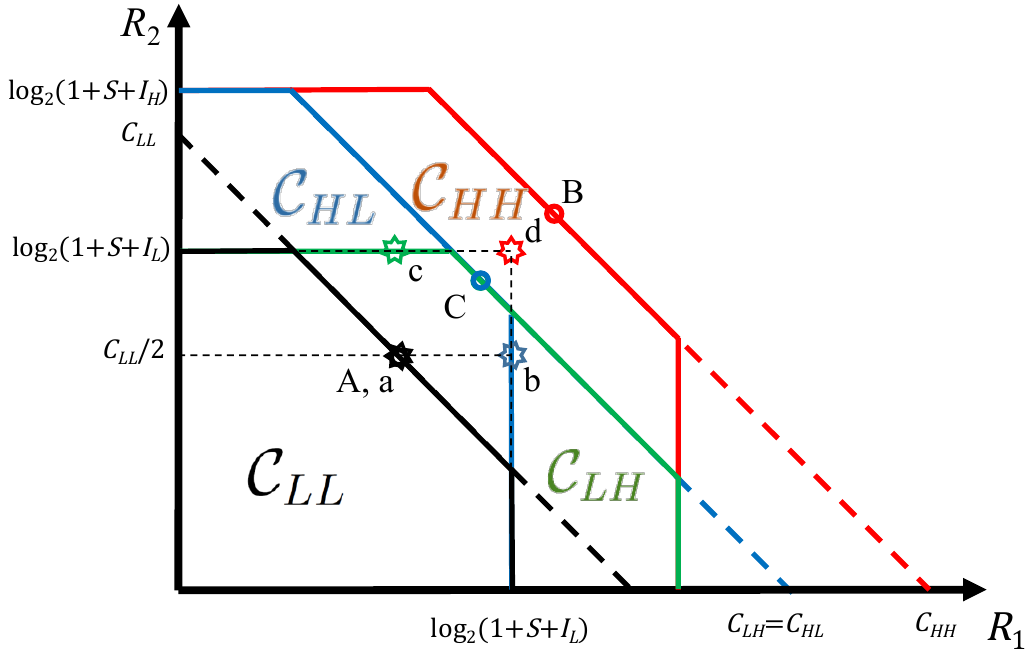}}\\
\subfigure[$2 \log_2(1+S+I_L) < C_{LH}$]{\label{Fig:Capacity_region_cc}\includegraphics[height=5.5cm]{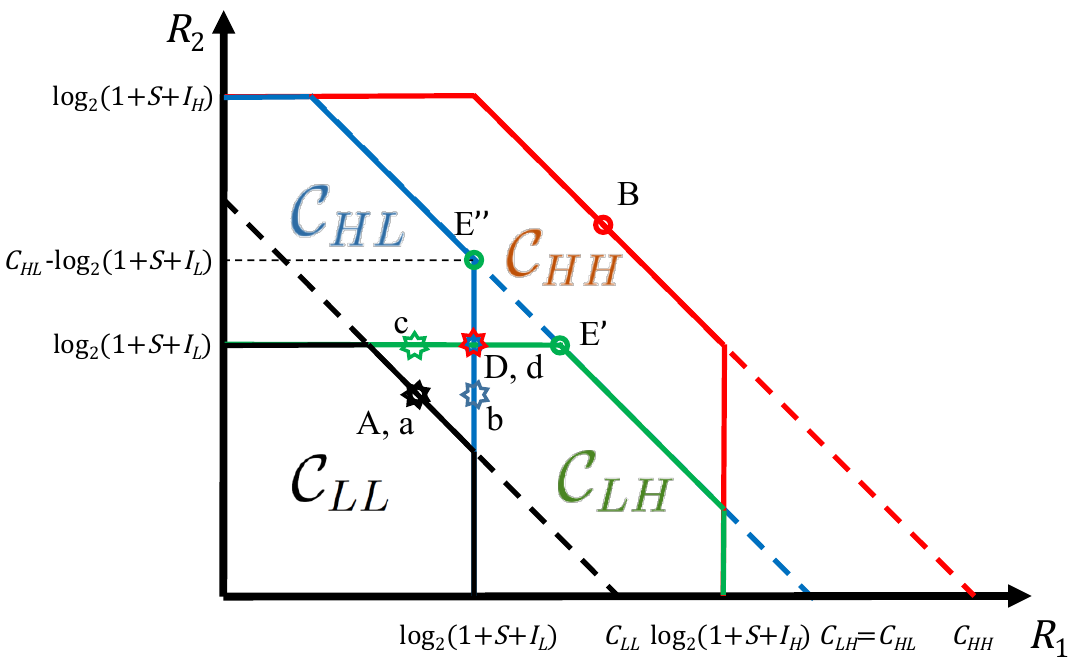}}
\caption{Capacity regions $\mathcal{C}_{xy}$ in (\ref{Capacity_region}) when the interference realizations are $I_1 = I_x$ and $I_2 = I_y$ if (a) $C_{LH} \le C_{LL}/2 + \log_2(1+S+I_L)$; (b) $C_{LL}/2 + \log_2(1+S+I_L) < C_{LH} \le 2 \log_2(1+S+I_L)$; and (c) $2 \log_2(1+S+I_L) < C_{LH}$. The points A, B, C, D, E, and E' denote the rates selected by the RCC in the C-RAN scheme discussed in Sec. \ref{Sec:C-RAN;Special} and the points a, b, c, and d indicate the four rate pairs $(R_x, R_y)$ selected by the F-RAN scheme in Sec. \ref{Sec:F-RAN;Special}.}
\label{Fig:Capacity_region_DC-CD}
\end{figure}

As discussed above, with C-RAN, the transmission rates $R_1 = R_1 ( I_1,  I_2)$, $R_2 = R_{2} ( I_1,  I_2)$ are selected by the RCC based on the available delayed CSI $\{I_1(t-d)= I_1, I_2(t-d)= I_2\}$ and joint data decoding is performed at the RCC. The set of achievable rate pairs $(R_1, R_2)$ with joint decoding at the RCC is given by the capacity region $\mathcal{C} ({I_1(t), I_2(t)})$ of the ergodic multiple access channel between the two users at the two RRSs. Using standard results in network information theory (see, e.g., \cite[Ch. 4]{GamalBook}), we have
\begin{equation} \label{Capacity_region}
\mathcal{C} ({I_1(t), I_2(t)}) = \left \{ 
\begin{array}{c|c}
& R_1 \le \log_2 (1 + S + I_2(t)) \\
(R_1, R_2) & R_2 \le \log_2 (1 + S + I_1(t)) \\
& R_1 + R_2 \le C ({I_1(t), I_2(t)}) \\
\end{array} \right \}.
\end{equation}
The capacity regions $\mathcal{C}_{LL}$, $\mathcal{C}_{LH}$, $\mathcal{C}_{HL}$, $\mathcal{C}_{HH}$ are illustrated in Fig. \ref{Fig:Capacity_region_DC-CD} under different conditions on the channels $(S, I_L, I_H)$. Observing the capacity regions in Fig. \ref{Fig:Capacity_region_DC-CD}, we note that, in the case $C_{LH} \le C_{LL}/2 + \log_2(1+S+I_L)$, there are achievable rate pairs that maximize the sum-rate in both capacity regions $\mathcal{C}_{LH}$ and $\mathcal{C}_{HL}$, namely the points marked as b and c in Fig. \ref{Fig:Capacity_region_aa}, while this is not true for case $C_{LH} > C_{LL}/2 + \log_2(1+S+I_L)$ as can be seen in Fig. \ref{Fig:Capacity_region_bb} and Fig. \ref{Fig:Capacity_region_cc}. This will play a role in the derivation of an achievable $\epsilon$-outage adaptive sum-rate below.

An outage occurs at time $t$ if the selected rate pair $(R_1, R_2)$ is outside the capacity region $\mathcal{C}_{I_1(t) I_2(t)}$. Accordingly, the outage probability in a time slot $t$ for which the CSI available at the RCC is $I_1 (t-d) =  I_1$ and $I_2 (t-d) =  I_2$ can be computed as 
\begin{equation} \label{C-RAN;Outage_Prob}
\textrm{Pr} [ (R_1, R_2 ) \notin \mathcal{C}({I_1(t), I_2(t)}) | I_1(t-d) =  I_1, I_2 (t-d) =  I_2 ].
\end{equation} 
An achievable $\epsilon$-outage adaptive sum-rate is summarized in the next lemma, where we defined the probabilities $P_{ x y}^{HH} = \beta^{H| x} (d) \beta^{H| y} (d)$, $P_{ x y}^{LH} = \beta^{L| x}(d) \beta^{H| y}(d)$, $P_{ x y}^{HL} = \beta^{H| x}(d) \beta^{L| y}(d)$, and $P_{ x y}^{LL} = \beta^{L| x}(d) \beta^{L| y}(d)$. The notation $P_{ x y}^{\bar x \bar y}$ indicates the probability of transitioning from delayed states $\{ I_1(t-d) = I_{ x}, I_2(t-d) =  I_{ y} \}$ to current states $\{ I_1(t)= I_{\bar x}, I_2(t) = I_{\bar y} \}$. 
\prop With C-RAN, an achievable $\epsilon$-outage adaptive sum-rate $R^{\text{C-RAN}} (d, \epsilon)$ is given as 
\begin{equation} \label{ASR_CC-CD}
R^{\text{C-RAN}} (d, \epsilon) = \pi_L^2 R_{LL} + 2 \pi_L \pi_H R_{LH} + \pi_H^2 R_{HH},
\end{equation}
with $R_{ x y}$ being defined as
\begin{equation} \label{ASR_CC-CD_xy}
R_{ x y} = \left \{ 
\begin{array}{lll}
C_{LL} & \hspace{0.3cm} \text{if} \,\,\,\, \epsilon \le P_{ x y}^{LL}, & \\
C_{LH} & \hspace{0.3cm} \text{if} \,\,\,\, P_{ x y}^{LL} < \epsilon \le 1- P_{ x y}^{HH}, & \\
C_{HH} & \hspace{0.3cm} \text{if}  \,\,\,\, 1- P_{ x y}^{HH} < \epsilon \le 1, &,
\end{array} \right.
\end{equation}
if $C_{LH} \le 2 \log_2 ( 1 + S + I_L)$, and as
\begin{equation} \label{ASR_CC-CD_xy}
R_{ x y}= \left \{ 
\begin{array}{lll}
C_{LL} & \hspace{0.3cm} \text{if} \,\,\,\, \epsilon \le P_{ x y}^{LL}, & \\
2 \log_2 ( 1 + S + I_L) & \hspace{0.3cm} \text{if} \,\,\,\, P_{ x y}^{LL} < \epsilon \le \widetilde P_{ x y}, & \\
C_{LH} & \hspace{0.3cm} \text{if} \,\,\,\, \widetilde P_{ x y} < \epsilon \le 1- P_{ x y}^{HH}, & \\
C_{HH} & \hspace{0.3cm} \text{if}  \,\,\,\, 1- P_{ x y}^{HH} < \epsilon \le 1, &
\end{array} \right.
\end{equation}
if $C_{LH} > 2 \log_2 ( 1 + S + I_L)$, with $\widetilde P_{ x y} = \min ( P_{ x y}^{LH}, P_{ x y}^{HL}) + P_{ x y}^{LL}$.
\begin{IEEEproof}
See Appendix \ref{Appendix:Prop1} for the proof. 
\end{IEEEproof}

\rem \label{OASR:D-RAN;nooutage}An outage event can be generally avoided only if transmitting always at the minimum sum-rate $C_{LL}$, since the latter yields rate pairs that are within the capacity region in all other states (see Fig. \ref{Fig:Capacity_region_DC-CD}). Therefore, with $d>0$ and $\epsilon=0$, the adaptive sum-rate of C-RAN is given by $R^{\text{C-RAN}} (d, 0) = C_{LL}$.

\rem In the absence of CSI delay, i.e., with $d=0$, the outage adaptive sum-rate $R^{\text{C-RAN}} (0, \epsilon)$ in (\ref{ASR_CC-CD}) with any $\epsilon \neq 1$ can be simplified as $R^{\text{C-RAN}} (0, \epsilon) =  \pi_L^2 C_{LL} + 2 \pi_L \pi_H C_{LH} + \pi_H^2 C_{HH}$.
\subsection{General Case}
We now consider the general case with multiple RRS/UE pair and multiple channel states. To this end, we define the following rate expression for any subset $\mathcal{L}=\{l_1, \dots, l_L \} \subseteq \mathcal{K}$ of RRSs
\begin{equation} \label{Capacity_General}
C (\{S_{j}\}_{j \in \mathcal{L}}, \{{\bf{I}}_j\}_{j \in \mathcal{L}}) \triangleq \mathbb{E} \left[\log_2 \det \left ( {\bf{I}} + {\bf{H}} (\{S_j\}_{j \in \mathcal{L}}, \{{\bf{I}}_j\}_{j \in \mathcal{L}}) {\bf{H}}^\dagger (\{S_j\}_{j \in \mathcal{L}}, \{{\bf{I}}_j\}_{j \in \mathcal{L}}) \right) \right],
\end{equation}
where we have introduced the channel matrix 
\begin{equation}
{\bf{H}} (\{S_j\}_{j \in \mathcal{L}}, \{{\bf{I}}_j\}_{j \in \mathcal{L}}) = [{\bf{h}}_1 (S_1, {\bf{I}}_1), \dots, {\bf{h}}_{l_L} (S_{l_L}, {\bf{I}}_{l_L}) ],
\end{equation} 
with ${\bf{h}}_j (S_j, {\bf{I}}_j) = [\sqrt{I_{jl_1}} e^{j \theta_{jl_1}}, \dots, \sqrt{S_{j}} e^{j \theta_{jj}},$ $ \dots, \sqrt{I_{j l_L}} e^{j \theta_{jl_L}} ]^T$. The expectation in (\ref{Capacity_General}) is taken over the random phases $\{ \theta_{ji} \}$ for $i, j \in \mathcal{L}$. Under joint data decoding at the RCC with full receiver CSI, when the CSI is $S_j (t)= S_{j}$ and $I_{ji}(t) = I_{ji}$, for all $i, j \in \mathcal{K}$ and $i \neq j$, the capacity region $\mathcal{C} ( \{S_j(t)\}, \{ I_{ji} (t)\} )$ of the ergodic multiple access channel between the UEs and the RCC is given as (see, e.g., \cite[Ch. 4]{GamalBook})
\begin{equation} \label{Capacity_region:General}
\mathcal{C} ( {\bf{S}}, {\bf{I}} ) = \left \{ (R_1, \dots, R_K): \sum_{j \in \mathcal{L}} R_j \le C (\{ S_j\}_{j \in \mathcal{L}}, \{ {\bf{I}}_j\}_{j \in \mathcal{L}}), \,\, \forall \mathcal{L} \subseteq \mathcal{K} \right \}.
\end{equation}
Note that (\ref{Capacity_region:General}) is an extension of (\ref{Capacity_region}) to the general scenario studied here.

When selecting the transmission rates for slot $t$, the RCC has delayed CSI, namely $\{S_j (t-d)\}$ and $\{I_{ji}(t-d)\}$ for $i, j \in \mathcal{K}$ and $i \neq j$, and is hence not informed about the current capacity region $\mathcal{C} ( \{S_j (t)\}, \{ I_{ji} (t)\} )$ in (\ref{Capacity_region:General}). To evaluate an achievable the $\epsilon$-outage adaptive sum-rate, we introduce an outage sum-rate region $\mathcal{C}^{\epsilon}  ( {\bf{S}}, {\bf{I}} )$ that has the property that, conditioned on $S_j (t-d) = S_j$ and $I_{ji} (t-d) = I_{ji}$, the set of rates $(R_1, \dots, R_K) \in \mathcal{C}^{\epsilon}  ( {\bf{S}}, {\bf{I}} )$ belong to the capacity region $\mathcal{C} ( \{S_j(t)\}, \{ I_{ji} (t)\} )$ with probability no smaller than $1 - \epsilon$. As a result of this definition, choosing a rate pair in $\mathcal{C}^{\epsilon}  ( \{S_j(t-d)\}, \{ I_{ji} (t-d)\} )$ guarantees a probability of outage smaller than or equal to $\epsilon$. 

We specifically propose to define
\begin{equation} \label{Outagerate_region:General}
\mathcal{C}^{\epsilon}_d (  {\bf{S}},  {\bf{I}} ) \triangleq \left \{ (R_1, \dots, R_K): \sum_{j \in \mathcal{L}} R_j \le C ( \{ F^{-1}_{S_j, d} (\bar \epsilon| S_j) \}_{j \in \mathcal{L}}, \{ F^{-1}_{I_{ji}, d} (\bar \epsilon| I_{ji}) \}_{j,i \in \mathcal{L}, i \neq j}), \,\, \forall \mathcal{L} \subseteq \mathcal{K} \right \},
\end{equation}
where $F^{-1}_{S_j, d} (\bar \epsilon| S_j)$ is defined as the state of $\bar \epsilon$-percentile of conditional distributions $\beta_S^{\centerdot | S_j}(d)$, that is, the maximum state value $x \in \{S_1, \dots, S_{N_S}\}$ such that $\textrm{Pr} [S_j(t) \le x | S_j (t-d) = S_{j}] \le \bar \epsilon$; $F^{-1}_{I_{ji}, d} (\bar \epsilon| I_{ji})$ is analogously defined as the state of $\bar \epsilon$-percentile of conditional distributions $\beta_I^{\centerdot| I_{ji}}(d)$; and $\bar \epsilon$ is the individual outage probability of each state such that $1 - ( 1 - \bar \epsilon)^{K^2} = \epsilon$ as in Proposition \ref{Prop:D-RAN}. 

The problem of maximizing the resulting $\epsilon$-outage adaptive sum-rate over the choice of the sum-rates $\{ R ( {\bf{S}}, {\bf{I}} )\}$ for $ {\bf{S}} \in \mathcal{N}_S^{K}$ and $ {\bf{I}} \in \mathcal{N}_I^{K-1 \times K}$ can be then formulated as
\begin{subequations} \label{OP_C-RAN}
\begin{eqnarray} 
R^{\text{C-RAN}} (d, \epsilon) = \underset { \{ R (  {\bf{S}},  {\bf{I}} ) \} \ge 0}{\textrm{maximize}} && \sum_{ {\bf{S}} \in \mathcal{N}_S^K} \sum_{ {\bf{I}} \in \mathcal{N}_I^{K-1 \times K}} \pi_{\bf{S}} \, \pi_{\bf{I}} \, R (  {\bf{S}},  {\bf{I}} ) \\
{\textrm{s.t.}} \hspace{0.5cm} && R (  {\bf{S}},  {\bf{I}} ) \in \mathcal{C}^{\epsilon}_d (  {\bf{S}},  {\bf{I}} ), \label{Constraint_OP_C-RAN}
\end{eqnarray}
\end{subequations}
where the constraint (\ref{Constraint_OP_C-RAN}) applies to all values $ {\bf{S}} \in \mathcal{N}_S^{K}$ and $ {\bf{I}} \in \mathcal{N}_I^{K-1 \times K}$. The problem (\ref{OP_C-RAN}) is a linear program (LP) and can be tackled using standard solvers. Note that as in Remark \ref{OASR:D-RAN;nooutage}, an outage event in case of the positive delay $d$ can be avoided if transmitting at the sum-rate $C ( \{ F^{-1}_{S_j, d} (\bar \epsilon| S_j) \}, \{ F^{-1}_{I_{ji}, d} (\bar \epsilon| I_{ji}) \})$.
\section{Fog radio access network (F-RAN)} \label{Sec:F-RAN}
With F-RAN, each RRS individually performs rate adaptation based on the available CSI, while the RCC performs joint data decoding in a centralized manner. 

\subsection{Analysis with two RRSs and UEs} \label{Sec:F-RAN;Special}
In this section, we consider the system model in Section \ref{Sec:C-RAN;Special} with two RRSs and UEs, constant direct channel over $T$ intervals, and two-state cross-channels. Accordingly, we will use the same notation introduced in Section \ref{Sec:C-RAN;Special} for the transition and stationary probabilities of the cross-channels as well as for the sum-rate $C (I_1, I_2)$ when the cross-channels equal $I_1(t)=I_1$ and $I_2(t)=I_2$.

With F-RAN, the transmission rate $R_j$ for user $j$ is selected by each RRS $j$ based on the available local CSI $I_j(t-d_e)$, which is subject to the scheduling delay $d_e$ as for D-RAN, while the RCC performs centralized joint data decoding on behalf of the RRSs. We define as $R_{L}$ and $R_{H}$ the rates selected by each RRS $j$ when $I_j(t)= I_L$ and $I_j(t)= I_H$, respectively. As for C-RAN, the outage probability is the probability that the rate pair $(R_1, R_2 )$ does not belong to the capacity region $\mathcal{C}_{I_1(t) I_2(t)}$. 

\prop With F-RAN, an achievable $\epsilon$-outage adaptive sum-rate $R^{\text{F-RAN}} (d_e, \epsilon)$ is given as 
\begin{equation} \label{ASR:F-RAN;Speical}
R^{\text{F-RAN}} (d_e, \epsilon) = 2 \pi_L^2 R_{L} (d_e, \bar \epsilon) + 2 \pi_L \pi_H ( R_{L} (d_e, \bar \epsilon) +  R_{H}(d_e, \bar \epsilon)) + 2 \pi_H^2  R_{H}(d_e, \bar \epsilon),
\end{equation}
where $\bar \epsilon = 1 - (1-\epsilon)^{1/K}$ and $R_{ x} (d_e, \bar \epsilon)$ is defined as 
\begin{equation} \label{ASR_DC-CD_L}
R_{x} (d_e, \bar \epsilon) = \left \{ 
\begin{array}{lll}
R_L & \hspace{0.3cm} \text{if} \,\,\,\, \bar \epsilon \le \beta^{L| x} (d_e), & \\
R_H & \hspace{0.3cm} \text{if} \,\,\,\, \beta^{L| x} (d_e) < \bar \epsilon \le 1, & 
\end{array} \right.
\end{equation}
with 
\begin{equation} \label{ASR:F-RAN;L}
R_{L} = \left \{ 
\begin{array}{lll}
C_{LL}/2 & \hspace{0.3cm} \text{if} \,\,\,\, C_{LH} > C_{LL}/2 + \log_2(1+S+I_L), & \\
C_{LL}/2 & \hspace{0.3cm} \text{if} \,\,\,\, C_{LH} \le C_{LL}/2 + \log_2(1+S+I_L) \,\,\, {\textrm{and}} \,\,\, \pi_L^2 > \pi_H^2, & \\
C_{LH} - \log_2(1+S+I_L) & \hspace{0.3cm} \text{if}  \,\,\,\, C_{LH} \le C_{LL}/2 + \log_2(1+S+I_L)\,\,\, {\textrm{and}} \,\,\, \pi_L^2 \le \pi_H^2, &
\end{array} \right.
\end{equation}
and
\begin{equation} \label{ASR:F-RAN;H}
R_{H} = \left \{ 
\begin{array}{lll}
\log_2(1+S+I_L) & \hspace{0.3cm} \text{if} \,\,\,\, C_{LH} > C_{LL}/2 + \log_2(1+S+I_L), & \\
C_{LH}-C_{LL}/2 & \hspace{0.3cm} \text{if} \,\,\,\, C_{LH} \le C_{LL}/2 + \log_2(1+S+I_L) \,\,\, {\textrm{and}} \,\,\, \pi_L^2 > \pi_H^2, & \\
\log_2(1+S+I_L) & \hspace{0.3cm} \text{if}  \,\,\,\, C_{LH} \le C_{LL}/2 + \log_2(1+S+I_L) \,\,\, {\textrm{and}} \,\,\, \pi_L^2 \le \pi_H^2. &
\end{array} \right.
\end{equation}
respectively.
\begin{IEEEproof}
As for Proposition \ref{Prop:D-RAN}, we impose that the individual outage probability for each UE-RRS pair be no larger than $\bar \epsilon = 1 - (1-\epsilon)^{1/K}$ so that the overall outage probability is no larger than $\epsilon$ by construction. To this end, we first evaluate the rates $R_{L}$ and $R_{H}$ selected to guarantee no outage, i.e., $\bar \epsilon = 0$, for each RRS $j$ when $I_j(t)= I_L$ and $I_j(t)= I_H$, respectively.

In order to guarantee no outage, the rate pair $(R_{L}, R_{L})$ must be inside the capacity region $\mathcal{C}_{LL}$, and hence, from Fig. \ref{Fig:Capacity_region_DC-CD}, the rate $R_{L}$ should be selected in the interval $[0,  C_{LL}/2]$. In a similar manner, the rate pair $(R_{L}, R_{H})$ (or $(R_{H}, R_{L})$) should be inside the capacity region $\mathcal{C}_{LH}$ (or $\mathcal{C}_{HL}$). Therefore, the rate $R_{H}$ can be no larger than $\min (C_{LH} - R_{L}, \log_2(1+S+I_L))$. Finally, the rate pair $(R_{H}, R_{H})$ must belong to the capacity region $\mathcal{C}_{HH}$, which is guaranteed by the conditions derived above. Based on these considerations, the adaptive sum-rate can be computed by solving the problem 
\begin{eqnarray} \label{OP_DC-CD}
\nonumber \underset {R_{L}}{\textrm{maximize}} && 2 (\pi_L^2 + \pi_L \pi_H) R_{L} + 2 (\pi_L \pi_H + \pi_H^2) \min \left (C_{LH} - R_{L}, \log_2(1+S+I_L)\right) \\
{\textrm{s.t.}} && 0 \le R_{L} \le C_{LL}/2,
\end{eqnarray}
where the objective is obtained by averaging the achievable rate. Solving the linear max-min program \cite{LinearMaxMinProblem} yields (\ref{ASR_DC-CD_L}).

Now, if the rate $R_{ x} (d_e, \bar \epsilon)$ in (\ref{ASR_DC-CD_L}) is selected by the RRS $j$, the individual outage probability for each UE $j$ does not exceed $\bar \epsilon$ by definition. In fact, with this choice, an outage occurs if the transmission rate $R_H$ for UE $j$ is chosen by each RRS $j$ when the local delayed CSI is $I_j(t-d_e)= I_{x}$ for $ x \in \{L, H\}$ and the current CSI is $I_j(t) = I_L$. This probability is equal to $\beta^{L|x} (d_e)$.
\end{IEEEproof}

\subsection{General Case}
With F-RAN, the transmission rate $R_j$ for user $j$ is selected by each RRS $j$ based on the available delayed CSI $S_j(t-d_e)$ and $I_{ji}(t-d_e)$ for $i \in \mathcal{K} \setminus \{j\}$ as for D-RAN, while the RCC performs centralized joint data decoding on behalf of the RRSs. The set of achievable rate pairs $(R_1, \dots, R_K)$ with joint decoding at the RCC when the channel states are $\{S_i (t)\}$ and $\{I_{ji}(t)\}$ is given by the capacity region $\mathcal{C} ( \{S_i(t)\}, \{ I_{ji} (t)\} )$ in (\ref{Capacity_region:General}). 

Based on the definition of outage sum-rate region in (\ref{Outagerate_region:General}), in the presence of the scheduling delay $d_e$, a probability of outage smaller than or equal to $\epsilon$ is guaranteed if the rate tuple $(R_1, \dots, R_K)$ is selected in $\mathcal{C}^{\epsilon}_{d_e} ( {\bf{S}}, {\bf{I}} )$ in (\ref{Outagerate_region:General}), when $S_j (t-d_e)= S_{j}$ and $I_{ji} (t-d_e)= I_{ji}$ for $i, j \in \mathcal{K}$ and $i \neq j$. The problem of maximizing the $\epsilon$-outage adaptive sum-rate over the choice of the rates $\{ R ( S_j,  {\bf{I}}_j) \}$ for $ S_j \in \mathcal{N}_S$ and $ {\bf{I}}_{j} \in \mathcal{N}_I^{K-1}$ under the outage sum-rate region $\mathcal{C}^{\epsilon}_{d_e} ( {\bf{S}},  {\bf{I}})$ can then be written as 
\begin{subequations} \label{OP_F-RAN}
\begin{eqnarray} 
R^{\text{F-RAN}} (d_e, \epsilon) = \underset { \{ R ( S_j, {\bf{I}}_j) \} \ge 0}{\textrm{maximize}} &&  \sum_{{\bf{S}} \in \mathcal{N}_S^K} \sum_{{\bf{I}} \in \mathcal{N}_I^{K-1 \times K}} \pi_{\bf{S}} \pi_{\bf{I}} \sum_{j \in \mathcal{K}} R ( S_j, {\bf{I}}_j) \\
{\textrm{s.t.}} \hspace{0.5cm} && (R ( S_1,  {\bf{I}}_1), \dots, R ( S_K, {\bf{I}}_K)) \in \mathcal{C}^{\epsilon}_{d_e} ( {\bf{S}}, {\bf{I}} ), \label{Constraint_OP_F-RAN}
\end{eqnarray}
\end{subequations}
where the constraint (\ref{Constraint_OP_F-RAN}) applies to all values ${\bf{S}} \in \mathcal{N}_S^K$ and ${\bf{I}} \in \mathcal{N}_I^{K-1 \times K}$. As for problem (\ref{OP_C-RAN}), problem (\ref{OP_F-RAN}) is an LP and can be solved using standard solvers. 

\section{Numerical results} \label{Sec:Numerical}
In this section, we evaluate the performance of D-RAN, C-RAN, and F-RAN in terms of the $\epsilon$-outage adaptive sum-rate as a function of key system parameters such as fronthaul and scheduling delays and mobile velocity. We set the carrier frequency to $1$ GHz ($c/\lambda = 1$ GHz with $c=3 \times 10^8 $ m/s) and the transmission interval duration to $T_p=0.1$ ms. Unless stated otherwise, we set the number of direct and cross channel states to $N_S=N_I=15$, the mobile velocity to $v=100$ km/h, and the SNRs of the desired and cross channel signal to $\gamma_S=5$ dB and $\gamma_I=0$ dB, respectively.

\begin{figure}[t!]
\centering
\includegraphics[width=13.5cm]{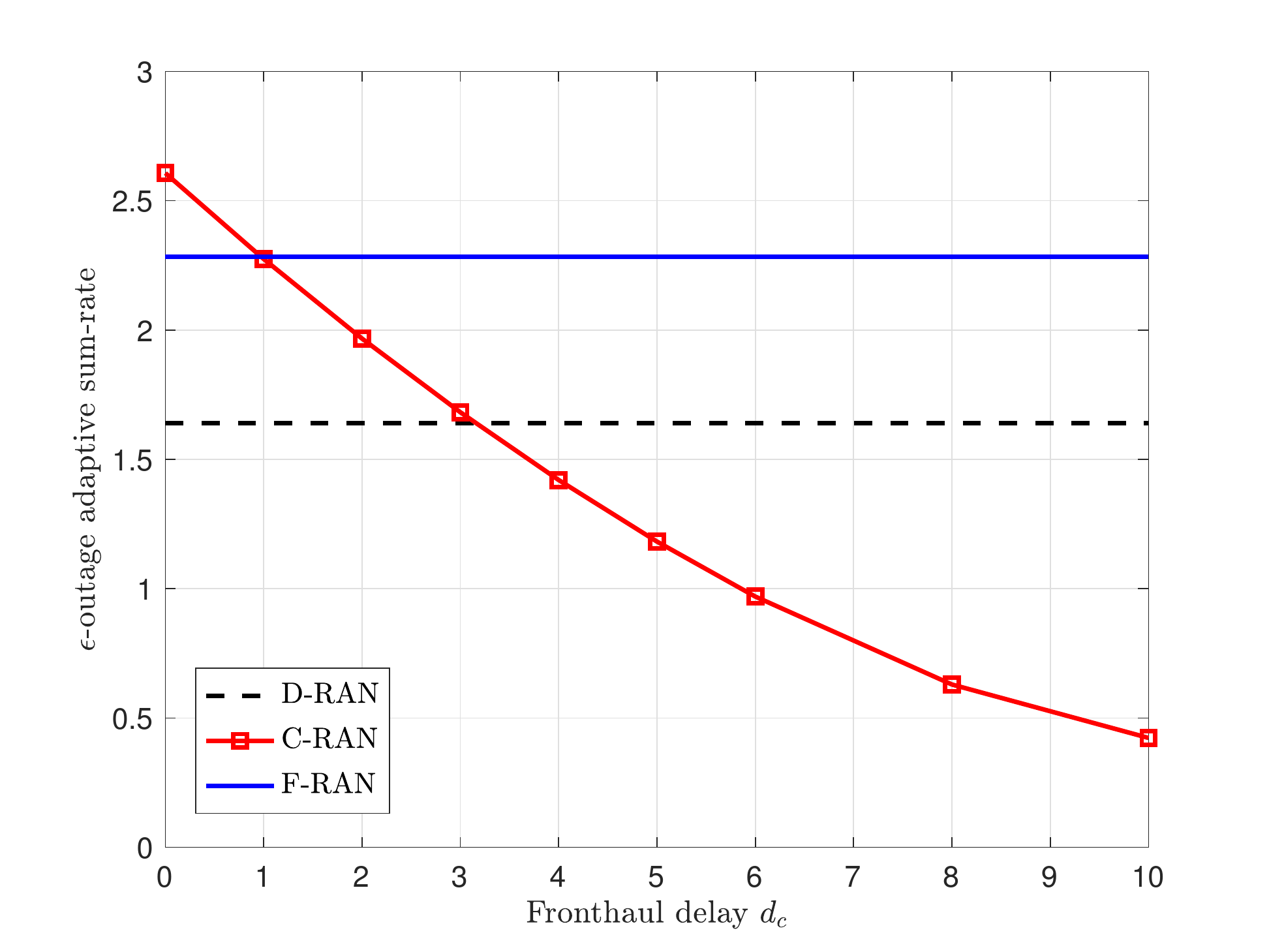}
\vspace{-0.5cm}
\caption{$\epsilon$-outage adaptive sum-rate vs. fronthaul delay $d_c$ for D-RAN, C-RAN, and F-RAN ($\epsilon = 0$, $d_e = 2$, $N_S=N_I=15$, $\gamma_S=5$ dB, $\gamma_I=0$ dB, and $v=100$ km/h).}
\label{Fig:GenerAdaptive sumrate_vs_dc}
\vspace{-0.5cm}
\end{figure}

We first investigate the impact of the fronthaul delay $d_c$ when the scheduling delay is $d_e = 2$ and the outage level is $\epsilon=0$. We recall that the fronthaul delay affects only C-RAN, whereby rate selection is performed based on CSI outdated by $d_e + d_c$ slots. From Fig. \ref{Fig:GenerAdaptive sumrate_vs_dc}, we observe that the centralized data decoding and control performed by C-RAN provides significantly performance gains over the decentralized data and control carried out by D-RAN, but only if the fronthaul delay $d_c$ is sufficiently small. Instead, when the fronthaul latency $d_c$ is large enough, the outdating of the CSI used by C-RAN to perform rate selection causes a significant performance degradation. This shows that centralized control based on global but delayed CSI can yield a degraded performance as compared to decentralized control based on local but more timely CSI. F-RAN is able to leverage the gains of centralized decoding of C-RAN, while at the same time also being robust to fronthaul delays thanks to decentralized control as in D-RAN.

\begin{figure}[t!]
\centering
\includegraphics[width=13.5cm]{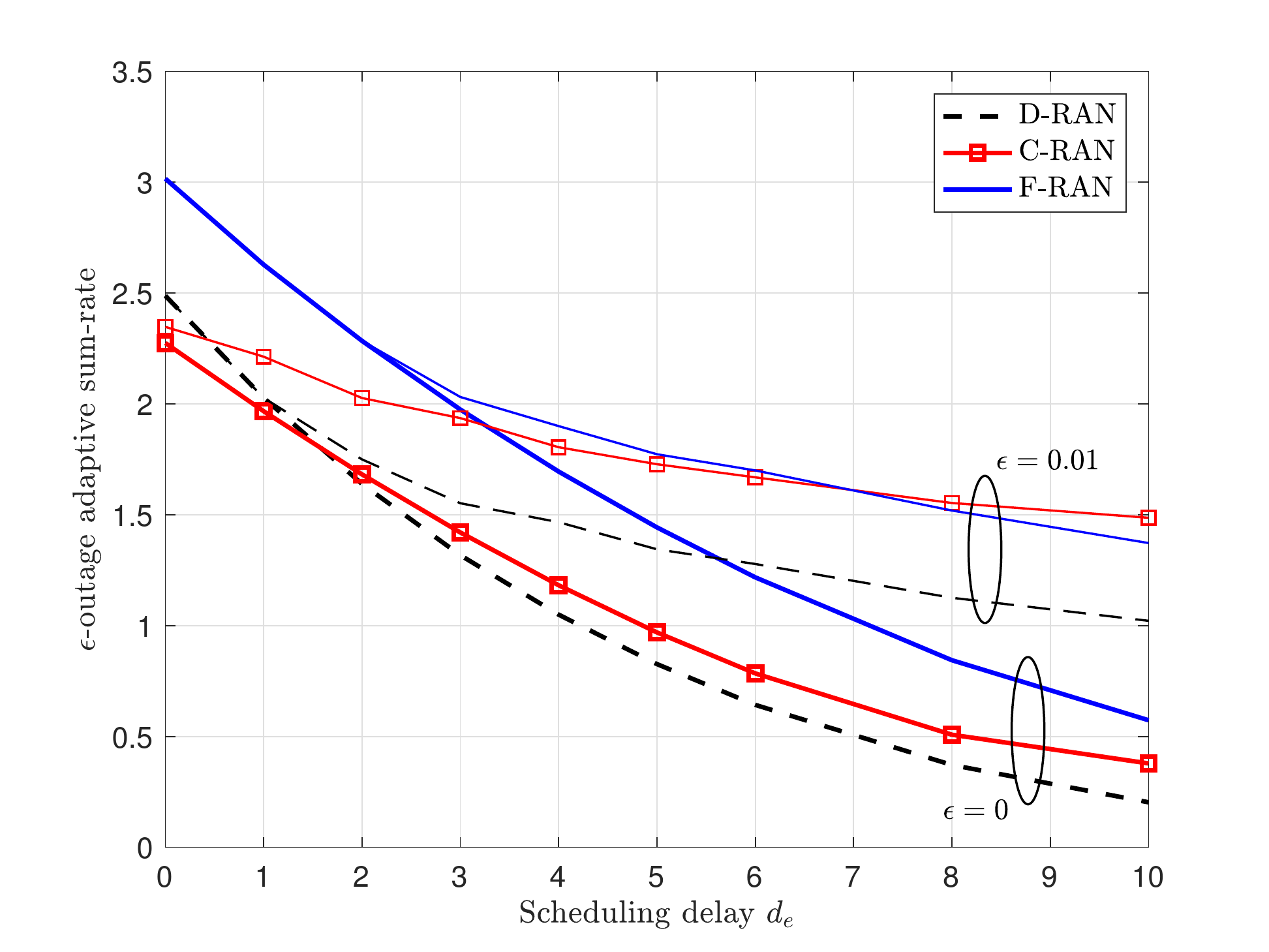}
\vspace{-0.5cm}
\caption{$\epsilon$-outage adaptive sum-rate vs. scheduling delay $d_e$ under the finite-state Markov model \cite{Wang95TVT} ($d_c=3$, $N_S=N_I=15$, $\gamma_S=5$ dB, $\gamma_I=0$ dB, and $v=100$ km/h).}
\label{Fig:GenerAdaptive sumrate_vs_de}
\vspace{-0.5cm}
\end{figure}

The impact of the scheduling delay $d_e$ is studied in Fig. \ref{Fig:GenerAdaptive sumrate_vs_de}, where the $\epsilon$-outage adaptive sum-rate is plotted versus $d_e$ with a fronthaul delay $d_c=3$. We recall that, while the fronthaul latency only affects the performance of C-RAN, rate selection for all schemes operates on an increasingly outdated CSI as $d_e$ becomes larger. We consider both $\epsilon = 0$ and $\epsilon = 0.001$. At the given value of $d_c$, F-RAN is seen to outperform both C-RAN and D-RAN for all values of $d_e$, with decreasing absolute gains as $d_e$ increases. We also see that C-RAN can outperform D-RAN for sufficiently large scheduling delay $d_e$, especially if one allows for a positive outage $\epsilon$. This is because the performance of decentralized control is degraded as $d_e$ grows larger.

\begin{figure}[h!]
\centering
\includegraphics[width=13.5cm]{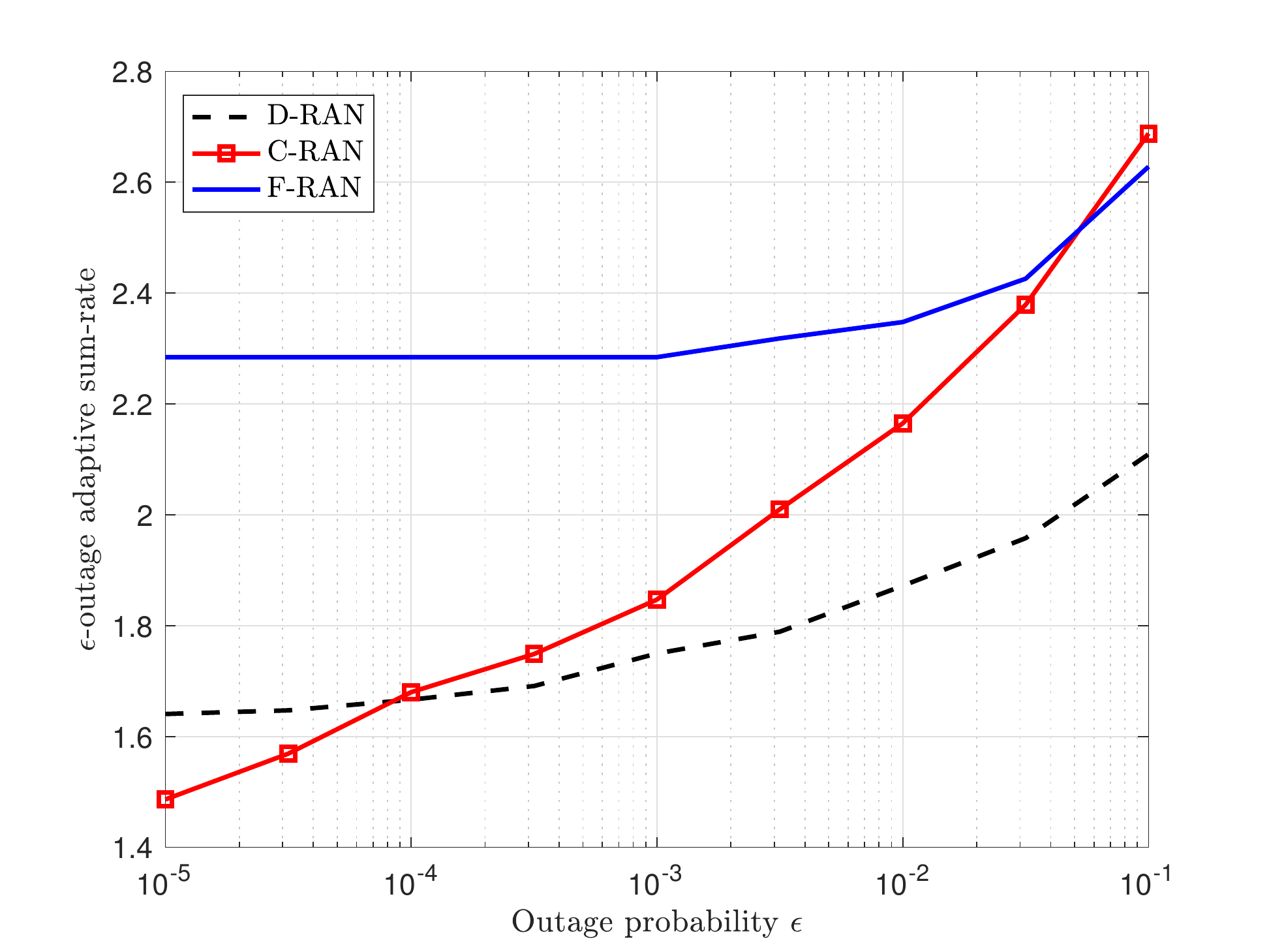}
\vspace{-0.5cm}
\caption{$\epsilon$-outage adaptive sum-rate vs. outage probability $\epsilon$ for D-RAN, C-RAN, and F-RAN ($d_c=5$, $d_e=2$, $N_S=N_I=15$, $\gamma_S=5$ dB, $\gamma_I=0$ dB, and $v=100$ km/h).}
\label{Fig:GenerAdaptive sumrate_vs_eps}
\vspace{-0.5cm}
\end{figure}

\begin{figure}[t!]
\centering
\includegraphics[width=13.5cm]{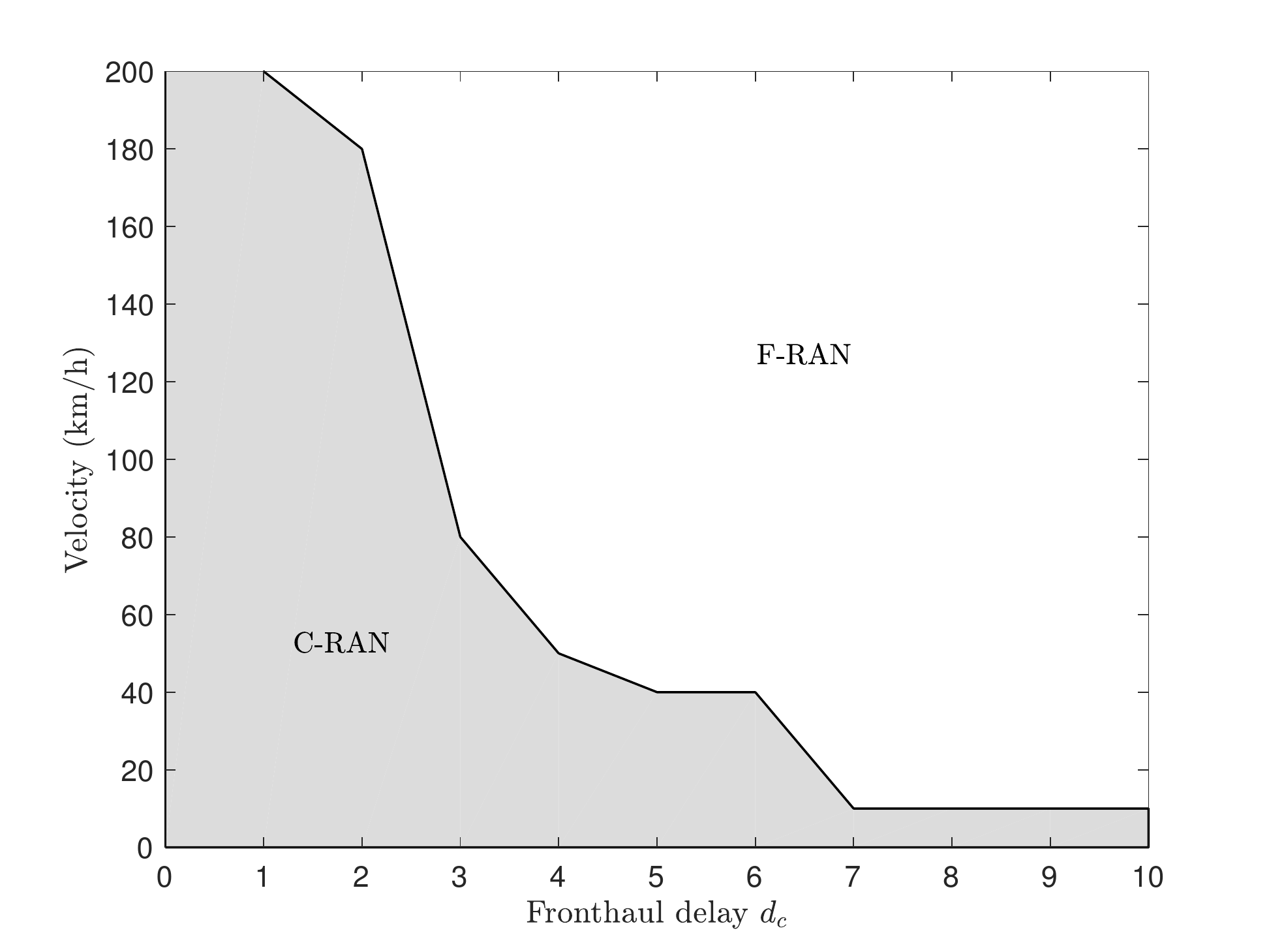}
\vspace{-0.5cm}
\caption{Regions of the plane $(d_c, v)$ in which F-RAN or C-RAN yield a larger $\epsilon$-outage adaptive sum-rate when allowing an outage of $\epsilon = 0.01$ ($d_e=3$, $N_S=N_I=12$, $\gamma_S=5$ dB, and $\gamma_I=0$ dB).}
\label{Fig:Comparison_region_vs_velNdc}
\vspace{-0.5cm}
\end{figure}

The impact of the outage level $\epsilon$ is further investigated in Fig. \ref{Fig:GenerAdaptive sumrate_vs_eps}, where we set $d_c=5$ and $d_e=2$. F-RAN is again seen to outperform both D-RAN and C-RAN, unless the allowed outage probability $\epsilon$ becomes large enough, here $\epsilon > 0.05$, in which case C-RAN can improve over F-RAN. In a similar way, if one accepts a sufficiently large outage probability, here $\epsilon > 0.0001$, C-RAN can perform better than D-RAN.


\begin{figure}[t!]
\centering
\includegraphics[width=13.5cm]{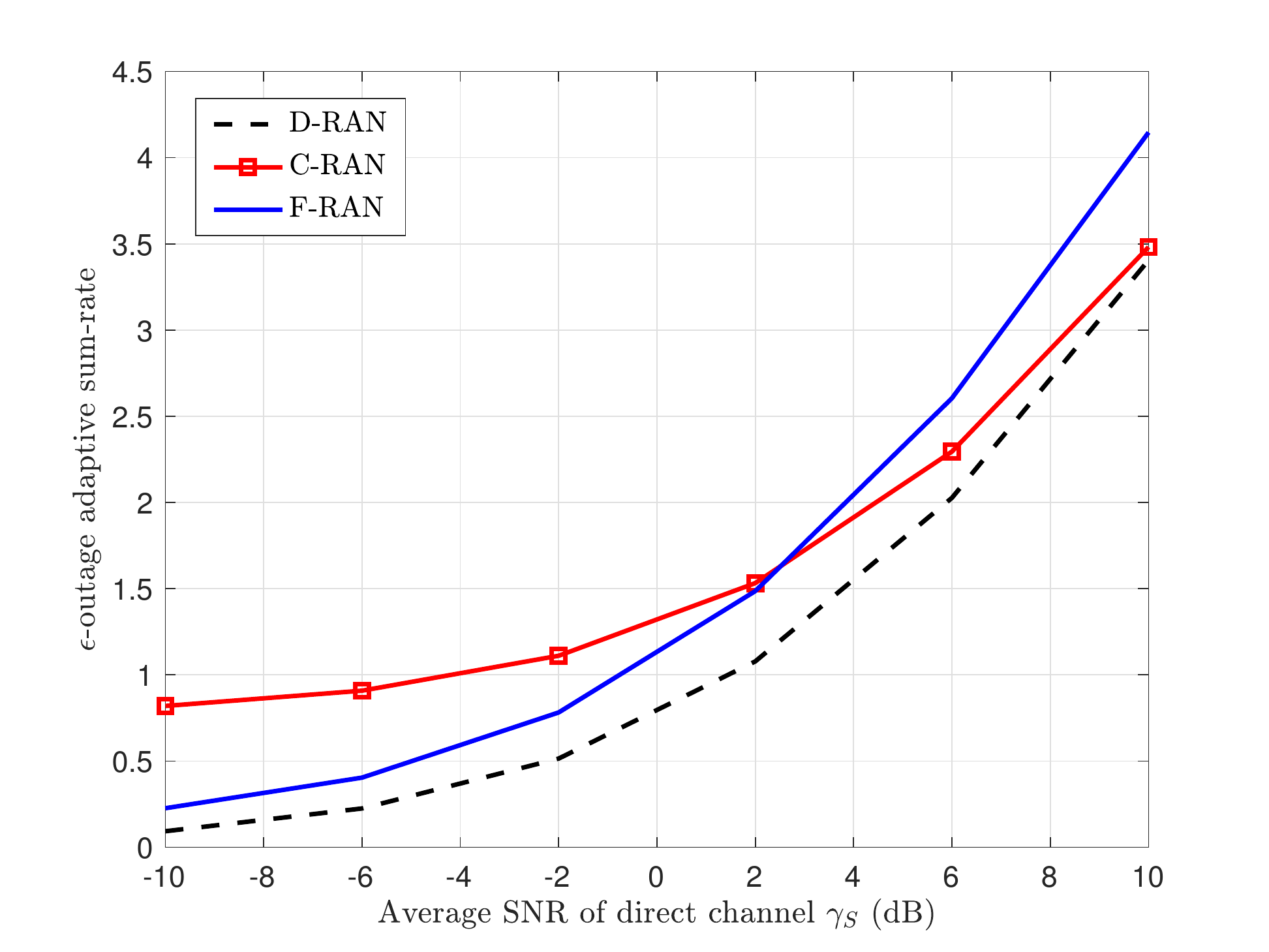}
\vspace{-0.5cm}
\caption{$\epsilon$-outage adaptive sum-rate vs. average SNR of direct channel states $\gamma_S$ for D-RAN, C-RAN, and F-RAN ($d_c=3$, $d_e = 2$, $\epsilon = 0.001$, $N_S=N_I=15$, $\gamma_I=0$ dB, and $v=100$ km/h).}
\label{Fig:GenerAdaptive sumrate_vs_GammaS}
\vspace{-0.5cm}
\end{figure}

\begin{figure}[t!]
\centering
\includegraphics[width=13.5cm]{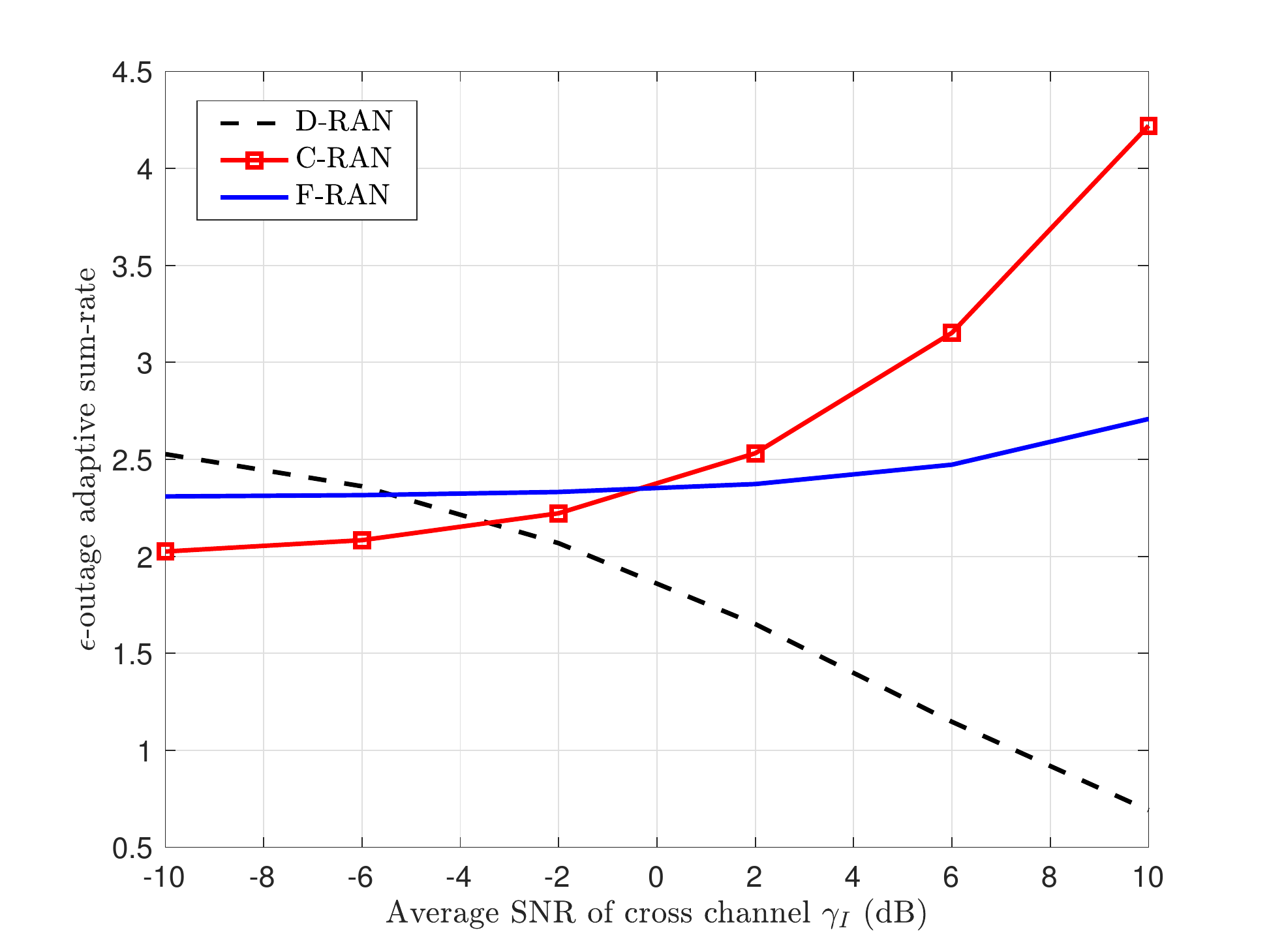}
\vspace{-0.5cm}
\caption{$\epsilon$-outage adaptive sum-rate vs. average SNR of interference channel states $\gamma_I$ for D-RAN, C-RAN, and F-RAN ($d_c=3$, $d_e = 2$, $\epsilon = 0.01$, $N_S=N_I=15$, $\gamma_S=5$ dB, and $v=100$ km/h).}
\label{Fig:GenerAdaptive sumrate_vs_GammaI}
\vspace{-0.5cm}
\end{figure}

In order to obtain additional quantitative insight into the operating regimes in which different functional splits are to be preferred, Fig. \ref{Fig:Comparison_region_vs_velNdc} shows the regions of the plane with coordinates given by the fronthaul delay $d_c$ and mobile velocity $v$ in which each scheme offers the best $\epsilon$-outage adaptive sum-rate. We set $d_e=3$, $N_S=N_I=12$, and $\epsilon = 0.01$. F-RAN is seen to be advantageous when the mobile velocity and the fronthaul delay are large enough. Note that F-RAN always outperforms D-RAN (not shown). The boundary line in Fig. \ref{Fig:Comparison_region_vs_velNdc} provides the maximum fronthaul delay $d_c$ that can be tolerated by C-RAN for a given value of the velocity $v$, while still yielding gains as compared to F-RAN (and hence also D-RAN).



Finally, in Fig. \ref{Fig:GenerAdaptive sumrate_vs_GammaS} and \ref{Fig:GenerAdaptive sumrate_vs_GammaI}, the $\epsilon$-outage adaptive sum-rate is plotted versus average SNR of direct and interference channel states, respectively, for $d_c=3$, $d_e=2$. It is seen that F-RAN is able to outperform C-RAN under the given conditions unless SNR $\gamma_S$ of direct channel is large or the average interfering channel gain $\gamma_I$ is small. This is because the centralized decoding performed by C-RAN is effective in compensating for low direct CSI channels by leveraging the cross-channel signal paths. In fact, C-RAN is able to treat the cross-channels as useful signals rather than as interference.


\section{Conclusions} \label{Sec:Conclusion}
The control-data separation architecture offers a promising guiding principle for the implementation of functional splits between edge and cloud, as enabled by NFV, in fog-aided 5G systems. In this paper, we have analyzed the relative merits of functional splits whereby rate selection and data decoding are carried out either at the edge or at the cloud by adopting the criterion of $\epsilon$-outage adaptive sum-rate. Among the main conclusions, this paper showed that the fully centralized architecture favored in the original instantiation of the C-RAN architecture is to be preferred only if the fronthaul latency is small or the time-variability of the channel is limited. Otherwise, a fog-based solution, whereby the control functionality of rate selection is carried out at the edge while joint data decoding is performed at the cloud, yields potentially significant gains. This conclusion demonstrates the value of decentralized but more timely CSI as compared to centralized but delayed CSI for the purpose of scheduling. 

Among interesting open problems, we mention here the study of models that allow for a more general definition of functional splits including a flexible demarcation line at the physical layer. Another relevant open aspect is the impact of outage events due to quasi-static fading, both in terms of coding strategies at the physical layer, such as the broadcast approach \cite{Shamai03TIT}, and of retransmission policies at the data link layer. Yet another issue is the modeling of both capacity and delays on the fronthaul links (see, e.g., \cite{Park14SPMAG}). Finally, it would be interesting to study downlink communication under the same assumptions on the heterogeneity of CSI available at edge and cloud considered in this paper. 

\appendices
\section{}\label{Appendix: Finite state Markov channel}
Here, we follow \cite{Wang95TVT} to define the $N_S$-state Markov model for the direct channel gains. Defining as $\gamma_S$ the average SNR of the direct channel states, the values $\{ S_1, \dots, S_{N_S} \}$ of the direct channel gains are obtained by selecting each value $S_m$ to be equal to the middle point in the quantization interval $[\Gamma_{S,m}, \Gamma_{S,m+1})$, which is identified by solving the equal-probability conditions $1/N_S = \exp (- \Gamma_{S,m}/ \gamma_S) - \exp (- \Gamma_{S,m+1}/ \gamma_S)$ with $\Gamma_{S,1} = 0$ and $\Gamma_{S, N_S+1} = \infty$ for $m = 1, \dots, N_S$. In a similar manner, defining as $\gamma_I$ the average SNR of the cross-channel states, the value $I_m$ is equal to the middle point in each quantization interval $[\Gamma_{I,m}, \Gamma_{I,m+1})$, which is obtained by solving the equations $1/N_I = \exp (- \Gamma_{I,m}/ \gamma_I) - \exp (- \Gamma_{I,m+1}/ \gamma_I)$ with $\Gamma_{I,1} = 0$ and $\Gamma_{I, N_I+1} = \infty$ for $m = 1, \dots, N_I$. For a mobile moving with velocity $v$ and transmitting with a carrier of wavelength $\lambda$, the transition probabilities are given as
\begin{equation} \label{TransitionProb_w_velocity}
p_{x, mn} = \left \{
\begin{array}{lll}
\frac{N(\Gamma_{x,m})T_p}{\pi_{x,n}} & \hspace{0.3cm} \text{if} \,\,\,\, m = n+1, & \\
\frac{N(\Gamma_{x,n})T_p}{\pi_{x,n}} & \hspace{0.3cm} \text{if} \,\,\,\, m = n-1, & \\
0 & \hspace{0.3cm} \text{if}  \,\,\,\, |m-n|>1, &
\end{array} \right.
\end{equation}
for $x \in \{S, I\}$, where $\Gamma_{x,m}$ is $N(\Gamma_{x,m}) = \sqrt{2 \pi \Gamma_{x,m} / \gamma_{x}} v/\lambda \exp (\Gamma_{x,m}/\gamma_{x})$ is the crossing rate of state $\Gamma_{x,m}$ for Clarke's model \cite{Wang95TVT} and $T_p$ is the duration of a transmission interval. 

\section{}\label{Appendix:Prop1}
The RCC chooses the rates $R_{1}$ and $R_{2}$ when $I_1 (t-d) = I_1$ and $I_2 (t-d) = I_2$ in such a way that the probability that the chosen rates are outside the capacity region $\mathcal{C}_{I_1(t) I_2(t)}$ for the current channel states $I_1 (t)$ and $I_2 (t)$ in (\ref{Capacity_region}) is less than $\epsilon$. Specifically, referring to Fig. \ref{Fig:Capacity_region_DC-CD} for an illustration, when $I_1 (t-d) = I_x$ and $I_2 (t-d) = I_y$: 
\begin{itemize}
\item If $\epsilon \le P_{xy}^{LL}$, the RCC selects $R_{1} = R_{2} = C_{LL}/2$ (point A in Fig. \ref{Fig:Capacity_region_DC-CD}); 
\item If $1 - P_{xy}^{HH} < \epsilon \le 1$, the RCC selects $R_{1} = R_{2} = C_{HH}/2$ (point B in Fig. \ref{Fig:Capacity_region_DC-CD}); 
\item If $P_{xy}^{LL} < \epsilon \le 1 - P_{xy}^{HH}$ and $C_{LH} \le 2 \log_2 ( 1 + S + I_L)$, the RCC selects $R_{1} = R_{2} = C_{LH}/2$ (point C in Fig. \ref{Fig:Capacity_region_aa} and Fig. \ref{Fig:Capacity_region_bb}); 
\item If $P_{xy}^{LL} < \epsilon \le \widetilde P_{xy}$ and $C_{LH} > 2 \log_2 ( 1 + S + I_L)$, the RCC selects $R_{1} = R_{2} = \log_2 (1 + S + I_L)$ (point D in Fig. \ref{Fig:Capacity_region_cc}); 
\item If $\widetilde P_{xy} < \epsilon \le 1 - P_{xy}^{HH}$ and $C_{LH} > 2 \log_2 ( 1 + S + I_L)$, the RCC selects either $R_{1} = C_{LH} - \log_2 (1 + S + I_L)$ and $R_{2} = \log_2 (1 + S + I_L)$ (point E$'$ in Fig. \ref{Fig:Capacity_region_cc}), or $R_{1} = \log_2 (1 + S + I_L)$ and $R_{2} =  C_{LH} - \log_2 (1 + S + I_L)$ (point E$''$ in Fig. \ref{Fig:Capacity_region_cc}), where the first rate pair is selected when $P_{xy}^{HL} + P_{xy}^{LL} < P_{xy}^{LH} + P_{xy}^{LL}$ and the other pair otherwise.
\end{itemize}
We will argue next that these choices guarantee a probability of outage (\ref{C-RAN;Outage_Prob}) no larger than $\epsilon$. 
\begin{enumerate}[A.]
\item When $R_{1} + R_{2} = C_{LL}$ (point A in Fig. \ref{Fig:Capacity_region_DC-CD}), the probability of outage can be easily seen to be zero, as discussed before, because the capacity region $\mathcal{C}_{LL}$ is included in a capacity region $\mathcal{C}_{I_1(t) I_2(t)}$ with any current channel states $\{ I_1(t), I_2(t) \}$. 
\item If the rates are selected so that $R_{1} + R_{2} = C_{HH}$ (point B in Fig. \ref{Fig:Capacity_region_DC-CD}), the upper bound (\ref{C-RAN;Outage_Prob}) on the outage probability is easily seen to be $1 - P_{xy}^{HH}$, which, in the relevant regime, does not exceed $\epsilon$, since any interference state other than $I_1(t) = I_H$ and $I_2(t) = I_H$ causes an outage. It can also be noted that the upper bound (\ref{C-RAN;Outage_Prob}) is in fact tight, since the outage events for the two users coincide.
\item If $C_{LH} \le 2 \log_2 ( 1 + S + I_L)$, the capacity regions (\ref{Capacity_region}) are shown in Fig. \ref{Fig:Capacity_region_aa} and Fig. \ref{Fig:Capacity_region_bb}. If the rates are selected to be $R_{1} = R_{2} = C_{LH}/2$ (point C in Fig. \ref{Fig:Capacity_region_aa} and Fig. \ref{Fig:Capacity_region_bb}), the upper bound on the probability of outage can be calculated as $P_{xy}^{LL}$, since only the interference state $\{ I_1(t) = I_L, I_2(t) = I_L \}$ causes an outage. Again, this probability is, by definition of the scheduling scheme, less than $\epsilon$, and the upper bound is in fact tight.
\item If $C_{LH} > 2 \log_2 ( 1 + S + I_L)$, the capacity regions (\ref{Capacity_region}) are shown in Fig. \ref{Fig:Capacity_region_cc}. If the rates are selected such that $R_{1} = R_{2} = \log_2 (1 + S + I_L)$ (point D in Fig. \ref{Fig:Capacity_region_cc}), the upper bound (\ref{C-RAN;Outage_Prob}) on the outage probability is easily seen to be tight and equal to $P_{xy}^{LL}$, which is smaller than $\epsilon$ in the relevant regime.
\item If $C_{LH} > 2 \log_2 ( 1 + S + I_L)$ and the rate pair $(R_{1}, R_{2}) = (C_{LH} - \log_2 (1 + S + I_L),\log_2 (1 + S + I_L))$ at E$'$ is selected, the upper bound (\ref{C-RAN;Outage_Prob}) on the probability of outage is equal to $P_{xy}^{HL}+P_{xy}^{LL}$ and tight. This is because an outage for both users is caused by the states $(I_1(t), I_2(t)) = (I_H, I_L)$ and $(I_1(t), I_2(t)) = (I_L, I_L)$. In a similar manner, if the rate pair $(R_{1}, R_{2}) = (\log_2 (1 + S + I_L), C_{LH} - \log_2 (1 + S + I_L))$ at E$''$ is selected, the probability of outage is given as $P_{xy}^{LH}+P_{xy}^{LL}$. Therefore, by selecting between the rate pairs at E$'$ and E$''$, we obtain the probability of outage $\widetilde P_{xy} = \min ( P_{xy}^{HL}+P_{xy}^{LL}, P_{xy}^{LH}+P_{xy}^{LL})$. This outage probability is also smaller than $\epsilon$ by construction of the scheduling scheme. 
\end{enumerate}

\bibliographystyle{IEEEtran}
\bibliography{refKJK}

\begin{thebibliography}{10}
\providecommand{\url}[1]{#1}
\csname url@samestyle\endcsname
\providecommand{\newblock}{\relax}
\providecommand{\bibinfo}[2]{#2}
\providecommand{\BIBentrySTDinterwordspacing}{\spaceskip=0pt\relax}
\providecommand{\BIBentryALTinterwordstretchfactor}{4}
\providecommand{\BIBentryALTinterwordspacing}{\spaceskip=\fontdimen2\font plus
\BIBentryALTinterwordstretchfactor\fontdimen3\font minus
  \fontdimen4\font\relax}
\providecommand{\BIBforeignlanguage}[2]{{%
\expandafter\ifx\csname l@#1\endcsname\relax
\typeout{** WARNING: IEEEtran.bst: No hyphenation pattern has been}%
\typeout{** loaded for the language `#1'. Using the pattern for}%
\typeout{** the default language instead.}%
\else
\language=\csname l@#1\endcsname
\fi
#2}}
\providecommand{\BIBdecl}{\relax}
\BIBdecl

\bibitem{Kang17WCNC}
J.~Kang, O.~Simeone, J.~Kang, and S.~Shamai, ``Control-data separation across
  edge and cloud for uplink communications in {C-RAN},'' \emph{Proc. IEEE
  Wireless Comm. and Net. Conf.}, pp. 1--6, San Francisco, CA, USA, Mar. 2017.

\bibitem{ChinaMobile13}
J.~Huang and R.~Duan, ``{C-RAN}: the road towards green {RAN},'' White Paper,
  ver. 3.0, China mobile Research Institute, Oct. 2013.

\bibitem{ChinaMobileNFGI}
J.~Huang and Y.~Yuan, ``White paper of next generation fronthaul interface,''
  \emph{[Online]. Available: labs.chinamobile.com/cran}, ver. 1.0, China mobile
  Research Institute, Jun. 2015.

\bibitem{5GNORMA}
{EU H2020 5G NORMA}, ``D3.2, {5G NORMA} network architecture {--} intermediate
  report,'' \emph{[Online] Available: https://5gnorma.5g-ppp.eu}, Jan. 2017.

\bibitem{Fettweis14SPMAG}
D.~Wubben, P.~Rost, J.~Bartelt, M.~Lalam, V.~Savin, M.~Gorgoglione, A.~Dekorsy,
  and G.~Fettweis, ``Benefits and impact of cloud computing on {5G} signal
  processing: Flexible centralization through cloud-{RAN},'' \emph{IEEE Sig.
  Proc. Mag.}, vol.~31, no.~6, pp. 35--44, Nov. 2014.

\bibitem{Simeone16JCN}
O.~Simeone, A.~Maeder, M.~Peng, O.~Sahin, and W.~Yu, ``Cloud radio access
  network: Virtualizing wireless access for dense heterogeneous networks,''
  \emph{Journal of Communications and Networks}, vol.~18, no.~2, pp. 135--149,
  Apr. 2016.

\bibitem{Rost17arXiv}
P.~Rost, C.~Mannweiler, D.~S. Michalopoulos, C.~Sartori, V.~Sciancalepore,
  N.~Sastry, O.~Holland, S.~Tayade, B.~Han, D.~Bega, D.~Aziz, and H.~Bakker,
  ``Network slicing to enable scalability and flexibility in {5G} mobile
  networks,'' \emph{arXiv:1704.02129v1}, Apr. 2017.

\bibitem{Chang16ICC}
C.-Y. Chang, R.~Schiavi, N.~Nikaein, T.~Spyropoulos, and C.~Bonnet, ``Impact of
  packetization and functional split on {C-RAN} fronthaul performance,''
  \emph{Proc. IEEE Int. Conf. on Comm.}, pp. 1--7, Kuala Lumpur, Malaysia, May
  2016.

\bibitem{Dotsch13Bell}
U.~Dotsch, M.~Doll, H.~P. Mayer, F.~Schaich, J.~Segel, and P.~Sehier,
  ``Quantitative analysis of split base station processing and determination of
  advantageous architectures for {LTE},'' \emph{Bell Labs Technical Journal},
  vol.~18, no.~1, pp. 105--128, Jun. 2013.

\bibitem{Rost2014WCL}
P.~Rost and A.~Prasad, ``Opportunistic hybrid {ARQ}: {E}nabler of
  centralized-{RAN} over nonideal backhaul,'' \emph{IEEE Wireless Comm. Lett.},
  vol.~3, no.~5, pp. 481--484, Jul. 2014.

\bibitem{Khalili16TETT}
S.~Khalili and O.~Simeone, ``Inter-layer per-mobile optimization of cloud
  mobile computing: A message passing approach,'' \emph{Transactions on
  Emerging Telecommunication Technologies}, vol.~27, no.~6, pp. 814--827, Feb.
  2016.

\bibitem{Tafazolli15CST}
A.~Mohamed, O.~Onireti, M.~Imran, A.~Imran, and R.~Tafazolli, ``Control-data
  separation architecture for cellular radio access networks: A survey and
  outlook,'' \emph{IEEE Communications Surveys and Tutorials}, vol.~18, no.~1,
  pp. 446--465, Jun. 2015.

\bibitem{Gulati16VTC}
S.~Gulati, B.~Natarajan, S.~Kalyanasundaram, and R.~Agrawal, ``Performance
  analysis of centralized {RAN} deployment with non-ideal fronthaul in
  {LTE}-advanced networks,'' \emph{Proc. IEEE Veh. Technol. Conf.}, pp. 1--5,
  May 2016.

\bibitem{Johnston15ISIT}
M.~Johnston and E.~Modiano, ``A new look at wireless scheduling with delayed
  information,'' \emph{Proc. IEEE Int. Symp. Info. Th.}, pp. 1407--1411, Hong
  Kong, Jun. 2015.

\bibitem{Sengupta16arXiv}
A.~Sengupta, R.~Tandon, and O.~Simeone, ``Fog-aided wireless networks for
  content delivery: Fundamental latency trade-offs,'' \emph{arXiv:1605.01690},
  May 2016.

\bibitem{SreekumarTIT15}
S.~Sreekumar, B.~Dey, and S.~Pillai, ``Distributed rate adaptation and power
  control in fading multiple access channels,'' \emph{IEEE Trans. Info. Th.},
  vol.~61, no.~10, pp. 5504--5524, Oct. 2015.

\bibitem{LTERel14Latency}
{3rd Generation Partnership Project (3GPP) TR 36.881}, ``Study on latency
  reduction techniques for {LTE} ({Release 14}),'' 2016.

\bibitem{NGMNOnline}
{NGMN Alliance}, ``Further study on critical {C-RAN} technologies,''
  \emph{[Online] Available: https://www.ngmn.org}, Mar. 2015.

\bibitem{Nikaein15CONF}
N.~Nikaein, ``Processing radio access network functions in the cloud: Critical
  issues and modeling,'' \emph{in Proc. of Int. Workshop on Mobile Cloud
  Computing and Services}, pp. 36--42, Paris, France, Sep. 2015.

\bibitem{Wang95TVT}
H.~S. Wang and N.~Moayeri, ``Finite-state {Markov} channel-a useful model for
  radio communication channels,'' \emph{IEEE Trans. on Veh. Technol.}, vol.~44,
  no.~1, pp. 163 --171, Feb. 1995.

\bibitem{Wei10TVT}
Y.~Wei, F.~R. Yu, and M.~Song, ``Distributed optimal relay selection in
  wireless cooperative networks with finite-state {Markov} channels,''
  \emph{IEEE Trans. on Veh. Technol.}, vol.~59, no.~5, pp. 2149--2158, Feb.
  2010.

\bibitem{Zheng13TWC}
K.~Zheng, F.~Liu, L.~Lei, C.~Lin, and Y.~Jiang, ``Stochastic performance
  analysis of a wireless finite-state {Markov} channel,'' \emph{IEEE Trans.
  Wireless Comm.}, vol.~12, no.~2, pp. 782--793, Jan. 2013.

\bibitem{MarkovChainBook}
{J. R. Norris}, \emph{Markov Chains}.\hskip 1em plus 0.5em minus 0.4em\relax
  Cambridge University Press, 1998.

\bibitem{GamalBook}
A.~E. Gamal and Y.-H. Kim, \emph{Network Information Theory}.\hskip 1em plus
  0.5em minus 0.4em\relax Cambridge University Press, 2011.

\bibitem{LinearMaxMinProblem}
J.~E. Falk, ``A linear max-min problem,'' \emph{Math. Program.}, vol.~5, no.~1,
  pp. 169--188, 1973.

\bibitem{Shamai03TIT}
S.~Shamai and A.~Steiner, ``A broadcast approach for a single-user slowly
  fading {MIMO} channel,'' \emph{IEEE Trans. Info. Th.}, vol.~49, no.~10, pp.
  2617--2635, Oct. 2003.

\bibitem{Park14SPMAG}
S.-H. Park, O.~Simeone, O.~Sahin, and S.~Shamai, ``Fronthaul compression for
  cloud radio access networks: signal processing advances inspired by network
  information theory,'' \emph{IEEE Sig. Proc. Mag.}, vol.~31, no.~6, pp.
  69--79, Nov. 2014.

\end{thebibliography}

\end{document}